\documentclass[%
 reprint,onecolumn,
 amsmath,amssymb,
 aps,
]{revtex4-2}
\usepackage{graphicx}
\usepackage{hyperref}
\usepackage{float}% Tim added: position figures/tables
\usepackage{dcolumn}   % needed for some tables
\usepackage{bm}        % for math
\usepackage{amssymb}   % for math
\usepackage{inputenc}
\usepackage{amsmath}
\usepackage{amssymb}
\usepackage{graphicx}
\usepackage[table, dvipsnames]{xcolor}
\usepackage{natbib}
\usepackage{mathrsfs}
\usepackage{color}
\usepackage{upgreek}
\newcommand{\meff}{m_{\mathrm{eff}}}
\newcommand{\rmd}{\mathrm{d}}
\newcommand{\xcrit}{x_{\mathrm{crit}}}
\newcommand{\Ecrit}{E_{\mathrm{crit}}}
\newcommand{\Elost}{E_{\mathrm{lost}}}
\newcommand{\Fes}{F_{\mathrm{es}}}

\newcommand{\beginsupplement}{%
        \setcounter{table}{0}
        \renewcommand{\thetable}{S\arabic{table}}%
        \setcounter{figure}{0}
        \renewcommand{\thefigure}{S\arabic{figure}}%
     \setcounter{section}{0}
        \renewcommand{\thesection}{S\arabic{section}}%
     }

\begin{document}

\title[Scalable nanomechanical logic gate]{Scalable nanomechanical logic gate}
\author{Erick Romero}
\thanks{These authors contributed equally to this work}
\author{Nicolas P. Mauranyapin}

\thanks{These authors contributed equally to this work}

\author{Timothy M. F. Hirsch}

\author{Rachpon Kalra}

\author{Christopher G. Baker}

\author{Glen I. Harris}

\author{Warwick P. Bowen}
%\auaddress{w.bowen@uq.edu.au}

\email{w.bowen@uq.edu.au}
\affiliation{The Australian Research Council Centre of Excellence for Engineered Quantum Systems, School of Mathematics and Physics, University of Queensland, St.Lucia, Queensland 4072, Australia}

\date{\today}

\begin{abstract}
    Nanomechanical computers promise robust, low energy information processing. However, to date, electronics have generally been required to interconnect gates, while no scalable, purely nanomechanical approach to computing has been achieved. Here, we demonstrate a nanomechanical logic gate in a scalable architecture. Our gate uses the bistability of a nonlinear mechanical resonator to define logical states. These states are efficiently coupled into and out of the gate via nanomechanical waveguides, which provide the mechanical equivalent of electrical wires. Crucially, the input and output states share the same spatiotemporal characteristics, so that the output of one gate can serve as the input for the next. Our architecture is CMOS compatible, while realistic miniaturisation could allow both gigahertz frequencies and an energy cost that approaches the fundamental Landauer limit. Together this presents a pathway towards large-scale nanomechanical computers, as well as neuromorphic networks able to simulate computationally hard problems and interacting many-body systems.
    \end{abstract}

\maketitle

\section{Introduction}

The miniaturisation of semiconductor electronics has fueled remarkable progress in computer performance over more than six decades. However, this progress has now markedly slowed~\cite{horowitz_11_2014,manipatruni_beyond_2018}. 

Thermal fluctuations of the energies of electrons and holes are a key barrier. Known as {\it Boltzmann's tyranny}~\cite{manipatruni_scalable_2019}, they enforce a minimum supply voltage for error-free computation which constrains the energy cost of logic operations above around 30 aJ \cite{manipatruni_scalable_2019,danowitz_cpu_2012,meindl_limits_2001}. This is a factor of $10^4$ higher than the  %Landauer limit, 
fundamental Landauer limit, which arises from the creation of entropy when information is erased in irreversible computing~\cite{landauer_irreversibility_1961,manipatruni_beyond_2018}.
Semiconductor electronics are also susceptible to degradation in harsh conditions, such as high temperature and radiation environments~\cite{senesky_harsh_2009,hassan_electronics_2018}. This imposes challenges for applications in space, medical and nuclear facilities, and in areas such as geothermal exploration and advanced propulsion systems~\cite{hassan_electronics_2018}. 

In nanomechanical computing, information is encoded in mechanical motion rather than electrical charge~\cite{roukes_mechanical_2004}. This evades
Boltzmann's tyranny, in principle allowing the Landauer limit to be reached~\cite{guerra_noise-assisted_2010}, and affords intrinsic robustness to radiation and temperature~\cite{lee_electromechanical_2010,song_additively_2019}. Nanomechanical computers also offer new computing modalities, such as adaptive information processing where the computer directly interacts with its environment via forces that it exerts or that are exerted upon it~\cite{yasuda_mechanical_2021}. These qualities, together with rapid advances in nanofabrication, have provided the impetus for much recent progress, including nanomechanical computers built from metamaterials~\cite{bilal_bistable_2017,song_additively_2019} and DNA-origami~\cite{yasuda_origami-based_2017, treml_origami_2018}, emulation of emergent phenomena such as
%the Ising model of 
ferromagnetism~\cite{mahboob_electromechanical_2016} and symmetry breaking~\cite{matheny_exotic_2019}, and reversible mechanical computing~\cite{wenzler_nanomechanical_2014}.
Nanomechanical elements are now even being applied as memories and interfaces for quantum computers~\cite{wollack_quantum_2022,von_lupke_parity_2022,chamberland_building_2022}. However, scaling nanomechanical gates into complex circuits remains an outstanding challenge. 

Unlike nanoelectromechanical computing, where information is stored in a combination of mechanical motion and electrical charge~\cite{ye_demonstration_2018,perrin_contact-free_2021}, scalable purely nanomechanical gates have yet to be developed. Purely mechanical gates generally rely on parametric interactions between mechanical waves~\cite{mahboob_interconnect-free_2011,hatanaka_broadband_2017,hatanaka_phonon_2013}, but these shift the frequencies of the bits so that the output of one gate cannot easily be used as the input of the next. Alternatively, direct physical contact has been used to interconnect gates~\cite{treml_origami_2018} but, lacking the equivalent of nanomechanical wires, this is constrained to simple circuit topographies.

Here, we report a highly efficient nanomechanical logic gate that is built into a scalable architecture. Our gate has a mechanical quality factor two orders of magnitude higher than previous nanomechanical gates~\cite{wenzler_nanomechanical_2014,hatanaka_broadband_2017,ilyas_cascadable_2019,guerra_noise-assisted_2010}. It is connected to nanomechanical waveguides which act as wires for the input and output bits. We show that input bits can be coupled into the gate with 99\% efficiency, and can drive transitions between two bistable states~\cite{hatanaka_phonon_2014, romero_propagation_2019}. This enables a new approach to nanomechanical computing~\cite{tadokoro_highly_2021}, for which we demonstrate a universal set of purely mechanical logic operations. Importantly -- and unlike previous approaches~\cite{mahboob_interconnect-free_2011,hatanaka_broadband_2017,hatanaka_phonon_2013} -- the output bits have the same frequency as the inputs and are therefore compatible with downstream gates. Our approach is CMOS compatible, facilitating both miniaturisation and scaling to complex architectures, and our modelling predicts that the Landauer limit is well within reach. This provides a pathway towards highly efficient, robust computer processing.

\section{Architecture}

\begin{figure}[h!]
     \centering
     \includegraphics[width=88mm]{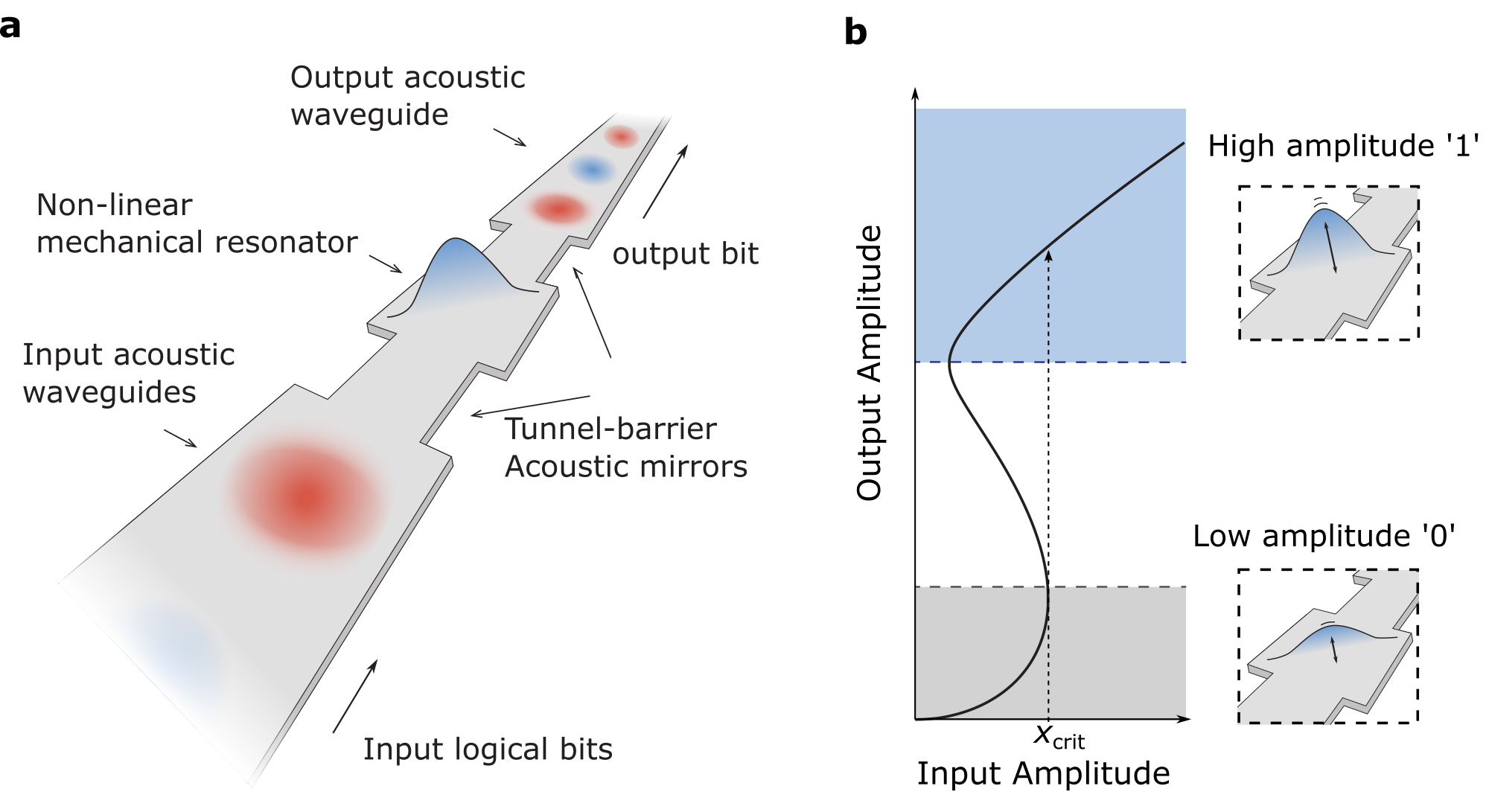}
         \caption{{\bf Concept for scalable nanomechanical logic.} {\bf a,} The nanomechanical gate is based on a nonlinear mechanical resonator that is created by placing two tunnel-barrier acoustic mirrors within a single mode acoustic waveguide. The gate can be precisely coupled to input and output waveguides by controlling the length of the tunnel-barriers. {\bf b,} Gate operations are implemented using the bistability of the resonator, which results in an abrupt jump from a `0' to a `1' state at a critical input amplitude $x_{\rm crit}$.
        }
     \label{fig:Concept}
\end{figure}

Our nanomechanical computing architecture is constructed on a silicon chip by combining single-mode acoustic waveguides~\cite{romero_propagation_2019} with evanescent tunnel barriers (Fig.~\ref{fig:Concept}{\bf a}). The tunnel barriers are created by locally narrowing the width of the waveguide until it no longer supports any acoustic modes. They then act as acoustic mirrors with reflectivity that can be finely tuned by controlling their length in fabrication~\cite{mauranyapin_tunneling_2021}. Here, we use pairs of closely-separated tunnel barriers to create a new kind of acoustic resonator reminiscent of an optical Fabry-P\'{e}rot cavity. These resonators form the nanomechanical logic gates in our computing architecture, with the single-mode acoustic waveguides used to input and output acoustic logical bits in the form of acoustic pulses.

Our acoustic resonators exhibit a strong hardening Duffing nonlinearity which introduces a bi-stability in the oscillation amplitude~\cite{schmid_fundamentals_2016}. This causes an abrupt transition from low to high amplitude when the input acoustic wave reaches a critical amplitude (Fig.~{\ref{fig:Concept} {\bf b}}). It has been suggested recently that this transition can be exploited for nanomechanical logic~\cite{tadokoro_highly_2021}, with the major advantage that the input and output bits have near identical spatiotemporal properties so that the output of one gate can be used as the input for the next. We follow this approach here, defining the low and high amplitude states, respectively, as the logical `0' and `1' states.

\section{Experiment}
\subsection{Fabrication}

\begin{figure}[h]
     \centering
     \includegraphics[width=178 mm]{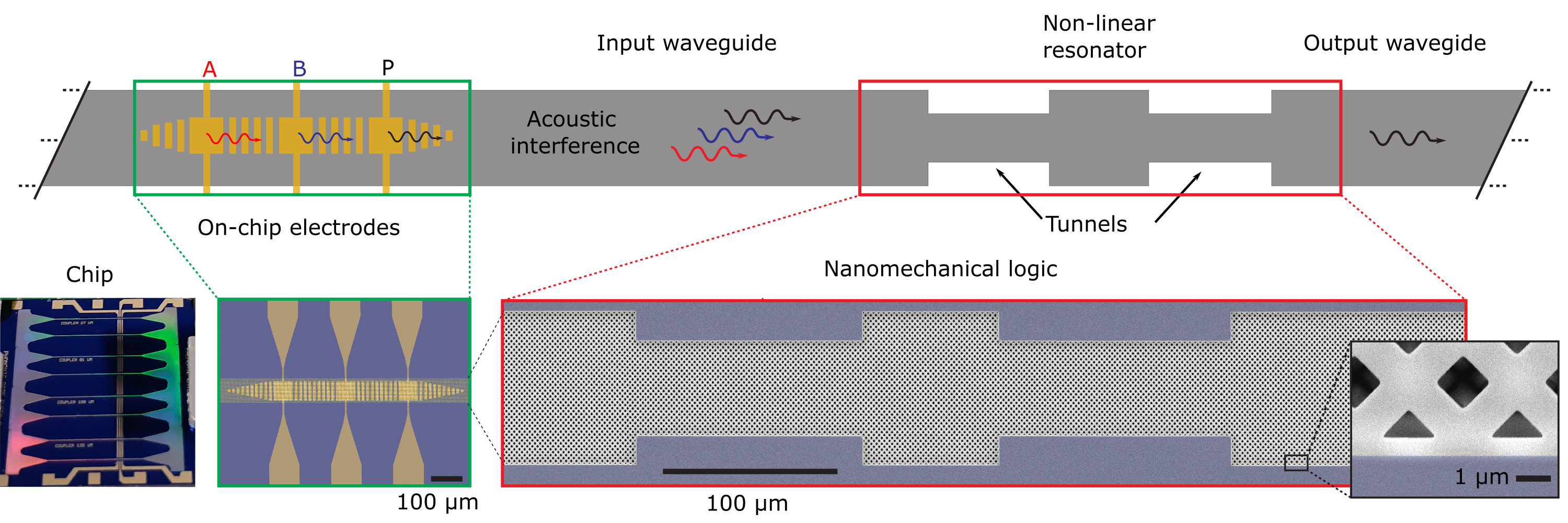}
     \caption{{\bf Fabricated nanomechanical computing architecture.} ({\it Top}) Schematic of the nanomechanical logic architecture, showing the on chip electrodes, acoustic interference, and nonlinear mechanical resonator. ({\it Bottom}) Microscope images of fabricated components. {\it Left:} Optical microscope image of device chip, showing eight logic devices. {\it Middle and right:} Scanning electron micrographs of the on-chip electrodes (false color) and nanomechanical resonator. To minimize acoustic impedance mismatch, and therefore reflections, gold strips are deposited both between and on either side of the electrodes with the length of the strips on the sides gradually tapered to zero to create an adiabatic transition to the bare acoustic waveguide (see Methods).  Inset: magnified image of the silicon nitride meshed structure.
               }
     \label{fig:Fabrication}
\end{figure}

The physical platform for our nanomechanical logical architecture uses a CMOS-compatible {\it mesh-phononic} approach that we have developed previously~\cite{mauranyapin_tunneling_2021}. As described in the Methods, here we fabricate an array of $80$~nm thick, 80~$\mu$m-square mechanical resonators from a stressed silicon nitride membrane~\cite{mauranyapin_tunneling_2021}. Each resonator is connected through evanescent tunnel barriers to single mode input and output waveguides. Figure~\ref{fig:Fabrication}, for example, shows a chip containing eight waveguide-coupled resonators each with different tunnel-barrier lengths. The factor-of-a-thousand aspect ratio between the size of a resonator and the thickness of the film results in a strong geometry-induced Duffing nonlinearity, reducing the energy required for logic operations (see Supplementary Information). Gold electrodes are patterned on the input waveguide. We use these to generate the acoustic inputs to each logic gate.

\subsection{Experimental Setup}

\begin{figure}
     \centering
     \includegraphics[width= 88mm]{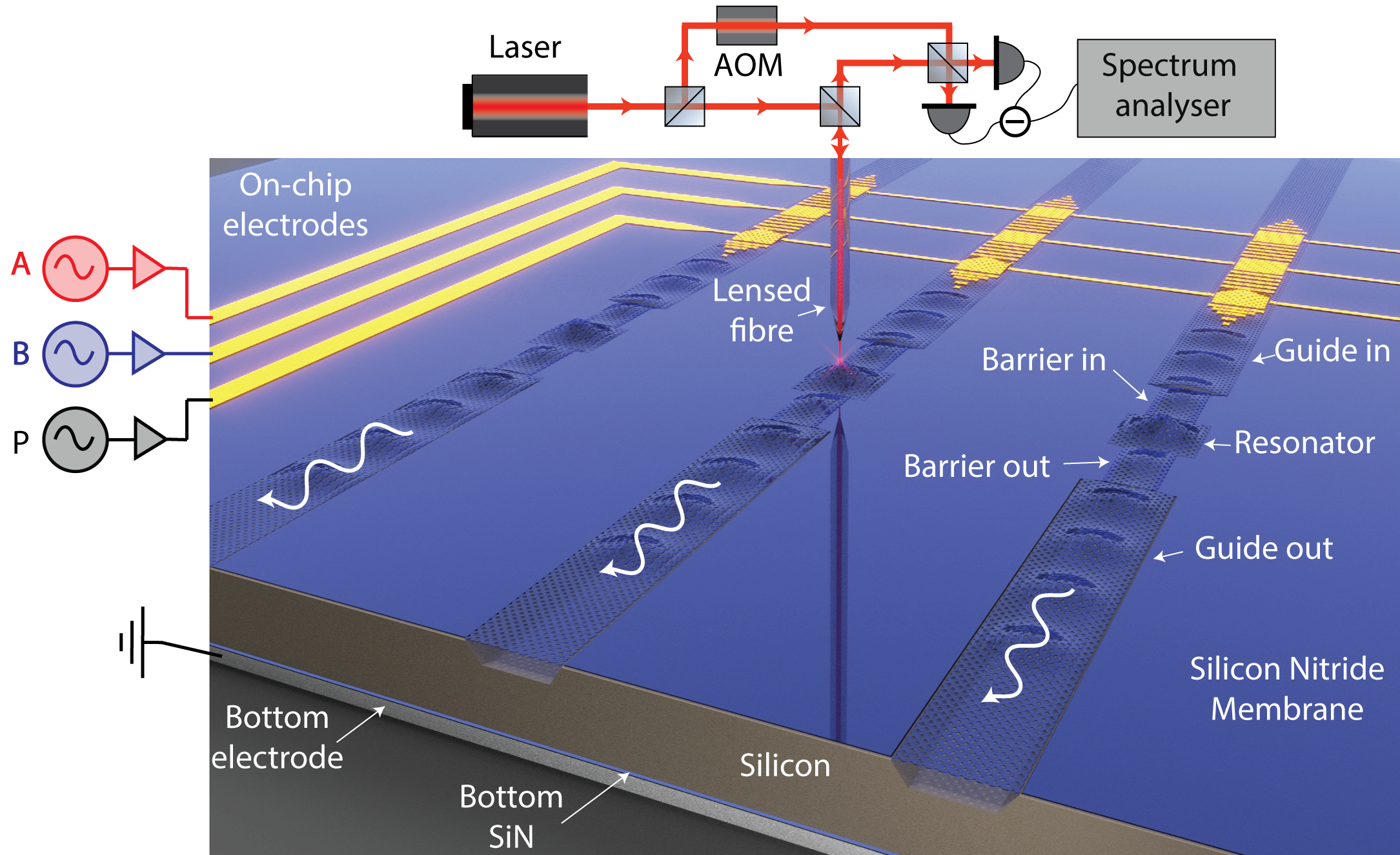}
     \caption{{\bf Experiment setup for acoustically driven nanomechanical logic.} Devices are actuated with electrostatics capacitive forces between three individual on chip electrodes A, B,P and a bottom electrode situated below the chip. Motion of the silicon nitride membrane is detected with a custom-built vibrometer combining a lensed fibre with heterodyne detection and recorded on a spectrum analyser. AOM: acousto-optic modulator.}
     \label{fig:Setup}
\end{figure}

The input acoustic logical bits to our nanomechanical logic gate, which we refer to as $A$ and $B$, are introduced by electrostatically actuating two of the electrodes situated on the input acoustic waveguide. The third electrode is used to provide a pump acoustic wave $P$ that drives the nanomechanical gate close to the nonlinear Duffing transition (see Fig.~\ref{fig:Setup}). This provides gain for logic operations, amplifying the amplitude of the output bit relative to the inputs and therefore offsetting any attenuation that might occur in propagation between logic gates. A combination of a DC voltage and a voltage oscillating at frequency $\Omega$, close to the natural resonance frequency of the mechanical resonator $\Omega_{\textnormal{m}}$, is applied between each electrode and a ground-plane under the chip (see Fig.~\ref{fig:Setup}). This results in three spatially separated forces on the waveguide $F_j = a_j \cos (\Omega t + \phi_j)$~\cite{romero_propagation_2019,mauranyapin_tunneling_2021}, where $j\in\{A, B, P \}$ label the electrodes that each force is applied to, and $a_j$ and $\phi_j$ are the amplitude and phase of the force. These forces generate out-of-plane acoustic waves that propagate along the input waveguide and interfere with each other to produce the combined acoustic input to the nanomechanical logic gate.

The electrodes and ground-plane are separated by the roughly half-millimetre thickness of the silicon chip. Since capacitive forces are roughly inversely proportional to the square of separation~\cite{bekker_free_2018}, this configuration would usually require kilovolts of potential difference to deflect the waveguide enough to access nonlinear oscillation amplitudes (see Supplemental Information). However, by choosing a boron-doped silicon substrate, which exhibits extremely high permittivity~\cite{cha_electrical_2018,cha_experimental_2018}, we are able to effectively displace the ground-plane potential to the top of the silicon substrate, around one micron away from the on-chip gold electrodes. This increases the capacitive forces by more than five orders of magnitude,  reducing the required potential difference to as little as tens of volts, as shown in the next section.

\newpage

\section{Results}
\subsection{Nanomechanical resonator characterization}

To detect the motion of the mechanical resonator, we use a custom-built laser vibrometer based on a combination of lensed fibre and heterodyne measurement~\cite{romero_propagation_2019,mauranyapin_tunneling_2021}, as described in the Methods. As a first test, we observe the thermal motion of seven nanomechanical resonators, each with different tunnel barrier lengths and therefore different quality factors. The input and output tunnel barriers lengths are chosen to be equal. In the absence of intrinsic dissipation, this forms an impedance matched acoustic cavity such that a resonant input acoustic wave is fully transmitted through the resonator and into the output waveguide.

\begin{figure}[h!]
     \centering
     \includegraphics[width=178mm]{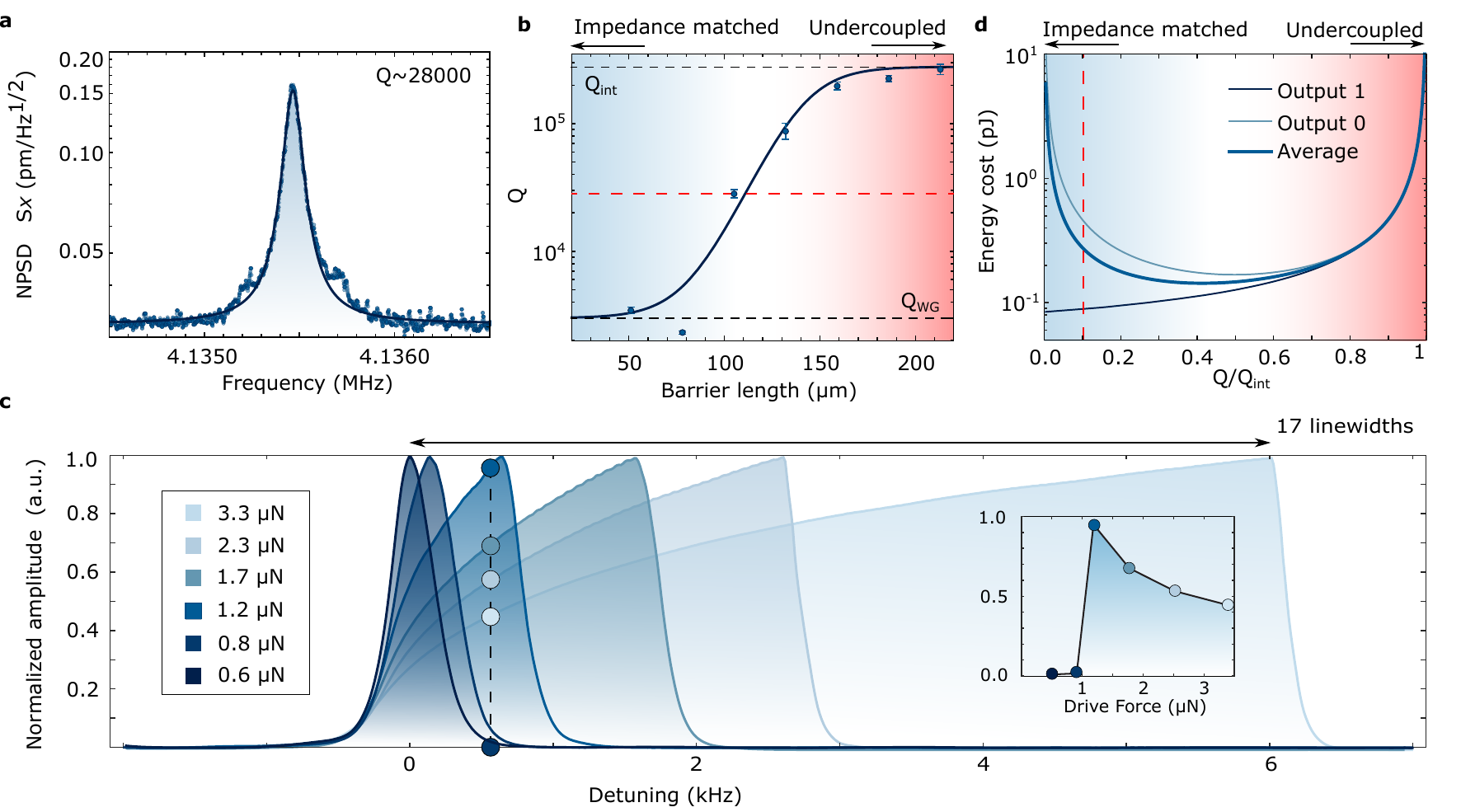}
     \caption{ {\bf Purely acoustic driving of nanomechanical logic gates.}  {\bf a,} Experimental measurements of the thermal noise power spectral density of a resonator with $Q\sim 28000$.  {\bf b,} Loaded quality factor as a function of the length of the acoustic mirrors.   {\bf c,} Normalized response of the nanomechanical meshed resonator to a swept-frequency acoustic tone. As the driving force is increased the response becomes increasing skewed, characteristic of a Duffing nonlinearity.  The inset shows the acoustic input amplitude for a 660~Hz detuning of the acoustic tone from the resonance frequency, illustrating the abrupt transition used for our logic operations.  {\bf d,} Predicted energy cost of logical operations as a function of loaded quality factor.  Dashed red line: predicted energy cost for our loaded quality factor of $Q=28000$.
   }
    \label{fig:QDuffing} 
\end{figure}

An example measurement of the power spectral density of a nanomechanical resonator with 105~$\mu$m barrier length is shown in Fig.~\ref{fig:QDuffing}{\bf a}. The mechanical resonance frequency of $\Omega_{\textnormal{m}}/2\pi=4.13$~MHz agrees well with our predictions based on finite-element modelling (see Supplementary Information.).

The ability to engineer the waveguide-resonator coupling enables wide tunability of both the resonator quality factor and transmission. Figure~\ref{fig:QDuffing}{\bf b} shows the loaded mechanical quality factors extracted from power spectra for each of the seven measured resonators. For barrier lengths above $150~\mu$m, the quality factor saturates at the intrinsic quality factor of the resonator, which we find to be $Q_{\rm int}=275,000$,  around two orders of magnitude higher than has previously been reported for a nanomechanical gate~\cite{wenzler_nanomechanical_2014,hatanaka_broadband_2017,ilyas_cascadable_2019,guerra_noise-assisted_2010}.
At barrier lengths below $50~\mu$m the quality factor instead approaches $Q_{\rm WG} \sim 3,000$, bounded above zero due to reflections at the ends of the waveguides. Apart from one resonator, for which a dust particle was observed on the meshed-membrane (barrier length of 78~$\mu$m), we find that the transition between the two regimes agrees well with modelling. 

A key development in our work is the ability to 
drive a nanomechanical resonator into the bi-stable regime through an acoustic waveguide. To explore this we drive the pump electrode $P$ on the input waveguide to introduce an acoustic wave. Sweeping the acoustic frequency through the nanomechanical resonance frequency from below, we monitor the resonator amplitude in response to the acoustic wave and observe the skewed Lorentzian response characteristic of a hardening Duffing resonator~\cite{brennan_jump-up_2008}. Figure~\ref{fig:QDuffing}{\bf c} shows the response of a resonator with a measured loaded quality factor of $Q=28,000$ ($105~\mu$m tunnel-barrier lengths)
for a range of drive forces.
We find that the bi-stable regime, where the resonance curve is multivalued, is reached for a combination of DC and root-mean-square AC drive voltages as low as $10$~V and $0.5$~V, respectively, corresponding to a drive force on the resonator of 1.7~$\mu$N. The inset of Fig.~\ref{fig:QDuffing}{\bf c} shows the normalised resonator amplitude as a function of drive strength at a fixed resonator-drive detuning of 660~Hz. At this detuning, we see a clear abrupt transition from low to high amplitude at a drive strength of 1.7~$\mu$N. In our computing scheme, these low and high amplitude states correspond to the logical `0' and `1' states.
 
Finite-element simulations provide a nonlinear Duffing coefficient of $\alpha=2.6\times10^{13}~\textnormal{N/m}^{3}$ for the resonator (see Supplementary Information.). From this we calculate a critical energy to reach the bistable regime required for computing of $E_{\mathrm{crit}}=k^2/(3 \alpha Q) = 28$~fJ~(see Supplementary Information), where $k=253$~N/m is the linear spring constant of the resonator. This roughly defines the minimum energy cost of a gate operation with our architecture, and is comparable to the energy cost of a gate in a 1995 Pentium Pro semiconductor microprocessor~\cite{zhirnov_minimum_2014}. In practice, the energy cost depends on whether the output is a `0' or a `1', since the lower amplitude of a `0' means the gate must reject more of the input energy in this case. It also depends on the duration of the logical pulses and amplitudes of the logical bits, among other factors. We model the energy cost
for both `0' and `1' outputs in the Supplementary Information using our experimental parameters (below). The results are shown as a function of loaded quality factor in Fig.~\ref{fig:QDuffing}{\bf d}, along with their average. The minimum average energy cost of 140~fJ is predicted to occur for a loaded quality factor of $Q \sim 0.41 \, Q_{\rm int}$. For technical reasons we choose to work with a device for which $Q \sim 0.1 \, Q_{\rm int}$ ($Q=28,000$). This results in a somewhat higher average energy cost of 270~fJ (dashed red line in  Fig.~\ref{fig:QDuffing}{\bf d}), but improved impedance matching -- on-resonant input acoustic waves are transmitted into the gate with 99\% efficiency and into the output waveguide with 81\% efficiency.

\subsection{Nanomechanical logic}

To demonstrate nanomechanical logic, we drive the $A$, $B$ and pump $P$ electrodes on the input waveguide with logic pulses. This creates spatially separated acoustic pulses that are synchronised to interfere with each other as they propagate, prior to the nanomechanical resonator (see Fig.~\ref{fig:Fabrication}). 
The pulses have a rectangular envelope, and the same carrier frequency $\Omega$. Careful choice of their phases and amplitudes enables a full set of Boolean logic gates to be implemented within the nanomechanical resonator. As an example, in Fig.~\ref{fig:ErrorRate}{\bf a} we demonstrate a universal NAND gate using a nanomechanical resonator with $Q\sim 28,000$. Here, the pump acoustic wave $P$ is chosen to have an amplitude that is high enough to drive the nanomechanical logic gate beyond the critical amplitude~$x_{\mathrm{crit}}=\sqrt{2\, k /(3 \,\alpha \,Q)}=15~$nm~(See Supplementary information), and into the `1' state with an oscillation amplitude of $\sim$18~nm. The $A$ and $B$ logical inputs are arranged to be out-of-phase with the pump, destructively interfering with it. Their amplitudes are chosen so that the nanomechanical logic gate transitions to the `0' state (with an oscillation amplitude below 2~nm) only if both inputs are `1's.

In Fig.~\ref{fig:ErrorRate}{\bf a}, we drive the gate with a sequence of all possible input combinations (`00', `01', `10' \& `11'), which allows us to record the truth table of the NAND gate in a single measurement of the peak of the nanomechanical power spectral density over time. As can be seen, the NAND gate operates as expected: the `00', `01' and `10' acoustic inputs all result in a `1' output state while a `11' input results in a `0' output state. Similar performance was achieved for  XOR, AND and NOR gates. From the measured amplitude of motion of the resonator, its quality factor and the resonator-drive detuning, our model gives an average gate energy cost of 270~fJ as described above. We have also shown faster gate operations, reducing the operation time from the 120~ms shown in  Fig.~\ref{fig:ErrorRate}{\bf a} to as short as 10~ms, slightly longer than the decay rate of the nanomechanical resonator. This reduced the average gate energy cost to 23~fJ.

\subsection{Error statistics and fatigue}

 To test for mechanical fatigue, we pumped a logic gate continuously for two months, corresponding to more than $10^{13}$ cycles, and separately performed around $10^6$ logic operations. We observed no statistically significant changes in mechanical quality factor or nonlinearity in either experiment.

\begin{figure}[h!]
     \centering
     \includegraphics[width=178 mm]{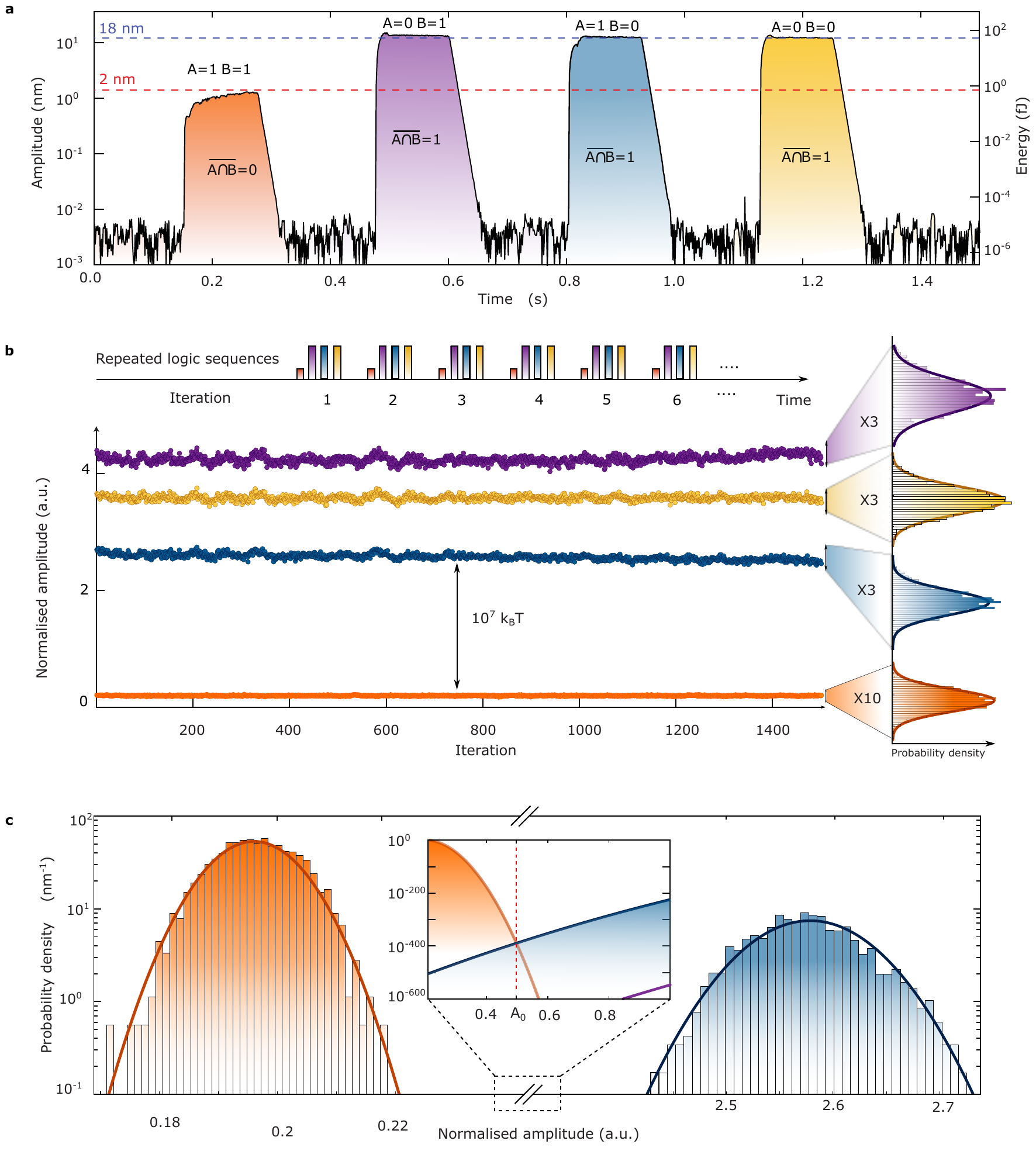}
     \caption{{\bf NAND gate.} {\bf a,} NAND gate truth table. Time trace showing the measured resonator amplitude for all four combinations of logical inputs when our nanomechanical device is configured to implement a universal NAND gate. 
 {\bf b,} Time evolution of the amplitudes of each of the four NAND gate outputs over 1500 measurements, corrected for slow time drift due to chip heating effects (see Supplementary Information). Histograms are plotted on the right. These are found to be consistent with Gaussian distributions (solid lines), as expected from the central limit theorem.
 {\bf c,} Histograms of the output bit amplitudes for `11' and `10' inputs. The inset shows an interpolation of the histogram fits, illustrating the low probability of a statistical error. All results used a drive-resonator detuning of approximately 660~Hz.
     }
     \label{fig:ErrorRate}
\end{figure}

To investigate the reliability of logical operations, we make repeated measurements of the NAND gate truth table of Fig.~\ref{fig:ErrorRate}{\bf a}. To do this, we automate the control and data taking systems as described in the Methods, repeating the sequence of four possible sets of input states (`00', `01', `10', \& `11') a total of 1500 times. We record the nanomechanical oscillation amplitude for each sets of input state, averaged over 10~ms. 
We observe no failures of gate operations for the full 6000 operations tested. 

To investigate the statistical probability of a failure, we correct for slow thermal drifts in the output amplitude and compute histograms of this normalised amplitude for each of the four sets of input states (see Supplementary information). 
As shown in Fig.~\ref{fig:ErrorRate}{\bf b}, the amplitude 
 fluctuations are far smaller than the separation of the `0' and `1' output states, consistent with our observation of no errors over the full sequence of measurements. Indeed, we find that the separation between the `0' bit value (red trace) and the lowest `1' bit value (blue trace) corresponds to
$\sim 10^7$ times the characteristic thermal energy ($k_B T$, with $k_B$ the Boltzmann constant and $T$ the temperature), so that thermally driven errors can be expected to be exceptionally rare. Fig.~\ref{fig:ErrorRate}{\bf c} confirms this, extrapolating the  histograms for these bit values to predict that it would take an 349$\sigma$ event for the `0' state to be mistaken for a `1', or a 47$\sigma$ event for the reverse to occur. Thus, statistical errors in logical operations using our nanomechanical computing architecture can be safely neglected, with errors likely dominated in practice by external perturbations to the system.

\section{Discussion}

The potential of nanomechanical technologies to allow robust, low energy computing has long been recognised~\cite{roukes_mechanical_2004}. However, the lack of a purely nanomechanical computing architecture that is scalable has been a major barrier, with electronics instead generally used to interconnect gates~\cite{ilyas_cascadable_2019}. The five essential criteria for scalable computing are~\cite{waser_nanoelectronics_2012,behin-aein_proposal_2010}: nonlinearity, gain, feedback prevention, functional completeness, and concatenability. While previous purely nanomechanical logic gates have, like ours, achieved functional completeness~\cite{wenzler_nanomechanical_2014}, used nonlinearities~\cite{hatanaka_broadband_2017,hafiz_microelectromechanical_2016}, and demonstrated gain~\cite{mahboob_phonon_2013}, none have been fully concatenatable due, either, to the reliance on physical contact between gates~\cite{treml_origami_2018}, or to differences in the frequencies of input and output bits~\cite{mahboob_interconnect-free_2011,hatanaka_broadband_2017,hatanaka_phonon_2014}. A similar challenge exists for many nanoelectromechanical gates, with reported part-in-ten-million conversion efficiencies between electronic and mechanical domains preventing efficient information transfer from one gate to the next~\cite{wenzler_nanomechanical_2014}.

Our nanomechanical logic architecture allows concatenation due to the use of both the Duffing bistability which ensures that input and output bits have the same spatiotemporal characteristics, and low-dissipation acoustic waveguides which ensure near-lossless transfer of information between gates. Indeed, the attenuation of our waveguides is well beneath a dB-per-cm~\cite{mauranyapin_tunneling_2021}, so that losses on transmission can be safely neglected. Our architecture also allows feedback prevention, so that acoustic reflections from one gate will not influence the function of upstream gates. For instance, higher-order mechanical resonances can be exploited to create a nanomechanical transistor for which the input (`gate') is naturally isolated from both pump (`source') and output (`drain'). We validate that this prevents feedback via  finite-difference time-domain simulations in the Supplementary  Information. Beyond conventional approaches to nanomechanical computing, we expect that this capacity to concatenate nonlinear mechanical resonators and control how they interact will open up new approaches to neuromorphic computing that overcome energy and density constraints of existing approaches using photonics, nanoelectromechanics, and semiconductor electronics~\cite{markovic_physics_2020}. It could also enable nonlinear phononic circuits that can simulate complex many-body phenomena such as symmetry breaking, synchronisation, and strongly-interacting spin systems~\cite{matheny_exotic_2019,fon_complex_2017,mahboob_electromechanical_2016}.

The energy cost of the nanomechanical logic operations we achieve is competitive with the best previously reported purely nanomechanical gates~\cite{hatanaka_broadband_2017}. However, at around $10^7 k_B T$ it remains far above the Landauer limit. This is caused by the need to pump the mechanical resonator to energies that exceed its critical energy of 28~fJ ($\sim 7 \times 10^6 \, k_B T$) in order to drive transitions between `0' and `1' states. 
%Speed is a further key factor for practical computation. 
Our demonstration is also slow, with each gate operation lasting at least 10~ms. Both speed and critical energy can be greatly improved through miniaturisation and material engineering. As shown in the Supplementary Information, reducing the width and thickness of the resonator to 2~$\mu$m and 10~nm, respectively, and decreasing the tensile stress by a factor of 100, would allow the Landauer limit to be reached and the speed of gate operations to be increased to around 50~kHz. Replacing the silicon nitride device layer with graphene would allow further miniaturisation into the nanoscale~\cite{cole_evanescent-field_2015,zhang_dynamically-enhanced_2020,jung_ghz_2019}, increasing the gate speed into the gigahertz regime while maintaining Landauer limited operation. %{\color{blue} \bf WPB: Need to add something about competitiveness with CMOS here once Erick has the numbers in.} 
Interestingly, in this regime, the mechanical resonance frequency is sufficiently high to allow passive cooling of the mechanical resonator into its quantum ground state using commercial cryogenic systems. Our computing architecture could also then be used as a coherent Ising machine to solve  non-deterministic polynomial-time (NP) hard computing problems~\cite{okawachi_demonstration_2020}, or potentially even as the basis for a purely nanomechanical quantum computer.

\section{Methods}

\subsection{Measurement setup}

To detect the mechanical motion of the membrane, we use a home-built laser vibrometer based on a combination of lensed fibre and heterodyne measurement~\cite{romero_propagation_2019,mauranyapin_tunneling_2021}. Light from a laser is reflected off the membrane (see Fig.~\ref{fig:Setup}). The periodic motion of the membrane at frequency $\Omega_{\textnormal{m}}$ modulates the phase of the reflected optical field. This modulation is then detected through its interference with a frequency-shifted local oscillator field $\Omega$ ~\cite{romero_propagation_2019,mauranyapin_tunneling_2021}. After detection, a beat-note that has an amplitude  proportional to the amplitude of motion of the membrane is observed using a spectrum analyser  (see Supplementary information). We find that the measurement is sufficiently precise to resolve both the thermal motion of the nanomechanical resonators that form our logical gates, and the driven motion of the input and output acoustic waveguides.

\subsection{Nanomechanical gate sequence}

To create the truth table sequence in Fig.~\ref{fig:ErrorRate}{\bf a}, we use a series of transistor-transistor-logic (TTL) pulses to trigger short bursts of coherent drive from three signal generators. The first of these three signal generators is used to create the electronic actuation for the pump, which consists of a sinusoidal wave at $\Omega$ gated by four 120~ms square pulses separated by 300~ms. The other two signal generators are used similarly to create the actuation for logical inputs A and B. The phase of the three sinusoidal signals are adjusted to create a NAND gate as explained in the the main text. The TTL pulses also trigger a spectrum analyser to record the amplitude of motion of the membrane at the drive frequency $\Omega$ with zero span and a resolution bandwidth of 510~Hz. In Fig.~\ref{fig:ErrorRate}{\bf a} we show a sequence that demonstrates a NAND gate, where the pump and two logic inputs $A$ and $B$ have a carrier frequency $\Omega$, which is slightly blue detuned from the bare resonant frequency $\Omega_{\textnormal{m}}$. The ratio of the amplitudes of the 0 and 1 states is close to a factor of ten, due to the nonlinear mechanical response of the resonator, as shown in Fig.~\ref{fig:QDuffing}{\bf c}.

\subsection{Fabrication details}

The devices are fabricated on a chip that is diced from a commercially available wafer which has a film of stoichiometric silicon-nitride (Si$_3$N$_4$) deposited on a silicon substrate. The Si$_3$N$_4$ film is $\sim$80~nm thick with an effective tensile stress of $\sigma_T$=0.67~GPa~\cite{mauranyapin_tunneling_2021}. We pattern the electrodes through a combination of electron beam lithography (EBL) and evaporation of gold ($\sim$50~nm), followed by lift-off. The hole pattern in the Si$_3$N$_4$ film is created using EBL and reactive-ion-etching. This pattern consists of a grid of 1~$\mu$m square holes separated by 2~$\mu$m (centre-to-centre). The Si$_3$N$_4$ membrane is then released from the silicon substrate via a potassium hydroxide wet etch. On each chip we define arrays of devices, each consisting of an 80~$\mu$m-square mechanical resonator connected with tunnel barriers to single mode input and output acoustic waveguides. Together, this procedure allows low-loss acoustic architectures to be built on a CMOS-compatible platform.

\subsection{Error statistic setup}

To record the data displayed in Fig.~\ref{fig:ErrorRate}{\bf b} and Fig.~\ref{fig:ErrorRate}{\bf c}, the pulse sequence shown in Fig.~\ref{fig:ErrorRate}{\bf a} is repeated over 1500 times. However, for this measurement, we operate with a resolution bandwidth of 100~Hz and only record the central point of each pulse displayed on the spectrum analyser.

\subsection{Adiabatic impedance transitions}
To avoid reflections and resonances between the three gold on-chip electrodes, the acoustic impedance between the silicon nitride waveguide without gold and the silicon nitride waveguide with gold needs to be matched. This is done by patterning nine gold pads on the waveguide between the on-chip electrodes, minimising the impedance difference. Similarly, to efficiently launch the acoustic wave in the membrane, adiabatic tapering of the impedance is used with ten decreasing size gold pads patterned on the left and right side of electrodes A and P (see Fig.~\ref{fig:Fabrication}).

%\backmatter

%\bmhead{Acknowledgments}
\subsection*{Acknowledgments}
This research was primarily funded by the Australian Research Council and the Lockheed Martin Corporation through the Australian Research Council Linkage Grant No. LP160101616. Support was also provided by the Australian Research Council Centre of Excellence for Engineered Quantum Systems (Grant No. CE170100009). This work was performed in part at the Queensland node of the Australian National Fabrication Facility and the Australian Microscopy \& Microanalysis Research Facility at the Centre for Microscopy and Microanalysis. G.I.H. and C.G.B. acknowledge their Australian Research Council (Grants No. DE210100848 and No. DE190100318), respectively. The authors would like to acknowledge HCPhotonics for providing a second harmonic module that was used in our heterodyne probing system.

%\bmhead{Author contributions}
\subsection*{Author contributions}

W.P.B., R. K. and C.B. conceived the idea. E. R. fabricated the devices. E.R., N.M. and T.M.F.H. carried out the experiment and data analysis with support from G.H. Theoretical modelling and simulations were done by N.M., E.R., and C.B. The manuscript was written by E. R., N. M and W.P.B with contributions from the rest of the authors. W.P.B. supervised the work.

\section*{Supplementary Information}
\beginsupplement

\section{Derivation of linear and nonlinear spring coefficients}
\label{sec:DerivationDuffing}
The origin of the non-linearity present in our system is known as geometric non-linearity~\cite{lifshitz_nonlinear_2008} with non-linear coefficient $\alpha$. This non-linearity emerges from a high amplitude displacement of the mechanical resonator. The calculation of $\alpha$ can be directly derived analysing the modeshape and displacement of our resonators. Here we derive $\alpha$ for the simplest system, a 1D string, extend the derivation for a 2D membrane using numerical simulations and finally derive it for a meshed membrane just as the one in our experiment. 

\subsection{String}
The simplest system to consider is a 1D string, with undeflected length 
\begin{equation}
    L=\int_0^L \rmd x, 
\end{equation}
while the length $L'$ of the deformed string is given by
\begin{equation}
    L'=\int_0^L (\Delta x^2 +\Delta u^2 )^{1/2}=\int_0^L \left(1 +\left(\frac{\partial u}{\partial x}\right)^2\right)^{1/2} \rmd x.
\end{equation}
In a deflected string, the local change in string length is 
\begin{equation}
  \rmd x\left(1 +\left(\frac{\partial u}{\partial x}\right)^2\right)^{1/2}-dx \simeq  \frac{1}{2} \left(\frac{\partial u}{\partial x}\right)^2 \rmd x, 
\end{equation}
which corresponds to a local strain
\begin{equation}
    \varepsilon(x)=\frac{1}{2} \left(\frac{\partial u}{\partial x}\right)^2.
    \label{eqstrain}
\end{equation}

The energy density $e(x)$ associated with this elongation is:
\begin{equation}
e(x)=\frac{1}{2} Y A  \varepsilon^2   = \frac{1}{8}A Y \left(\frac{\partial u}{\partial x}\right)^4,
\end{equation}
where $A$ is the cross-sectional area of the string, and $Y$ the Young's modulus of the material. The total elongation energy is thus given by the integral over the volume
\begin{equation}
E_{\mathrm{elongation}}=\int_0^L e(x) \rmd x=\frac{1}{8} Y A \int_0^L \left(\frac{\partial u}{\partial x}\right)^4 \rmd x.
\end{equation}
Identifying this with a quartic Duffing potential energy of the form $E=\frac{1}{4}\alpha x^4$, we get $\alpha$ the non-linear coefficient
\begin{equation}
\boxed{    \alpha=\frac{1}{2} Y A \int_0^L \left(\frac{\partial u_0}{\partial x}\right)^4 \rmd x},
    \label{Eqalphastring}
\end{equation}
where $u_0$ is the unperturbed normalized displacement profile, normalized such that $\mathrm{max}(u_0)=1$. 

Similarly, since the energy stored in the string's kinetic energy/tension is given by
\begin{equation}
    E=\frac{1}{2} \rho A \int_0^L \left(\frac{\partial u }{\partial t}\right)^2 \rmd x= \frac{1}{2}\sigma A\int_0^L \left(\frac{\partial u}{\partial x}\right)^2 \rmd x
\end{equation}
we can identify the linear spring $k$ and the effective mass $m_{\mathrm{eff}}=k/\Omega^2$,

\begin{equation} \label{eq:kstring}
    k=\rho \Omega_{\textnormal{m}}^2 A \int u_0^2(x) \rmd x= \sigma A\int_0^L \left(\frac{\partial u_0}{\partial x}\right)^2 \rmd x, \hspace{0.5 cm}\textnormal{and} \hspace{0.5 cm} m_{\mathrm{eff}}=\rho A \int u_0^2(x) \rmd x, 
\end{equation}
both in geometric terms.

\subsection{Membrane}
For the 2D case, in the regime that the resonator is thin (i.e. the resonator is thin enough to behave as a membrane and not a plate), the bending stiffness is neglected, and the Young's modulus is assumed isotropic. The membrane's Duffing nonlinear coefficient $\alpha$, by analogy to the string case (Eq. \eqref{Eqalphastring}), takes the form: 
\begin{equation}    \label{eq:alphamembrane}
\boxed{
    \alpha=\frac{1}{2} Y\, h \iint\left( \left( \frac{\partial u_0}{\partial x}\right)^4 + \left(\frac{\partial u_0}{\partial y}\right)^4\right) \rmd x\, \rmd y},
\end{equation}
where $h$ is the membrane thickness. Similarly, the spring constant $k$ and effective mass $m_{\mathrm{eff}}=k/\Omega_{\textnormal{m}}^2$ take the form
\begin{equation}
    k
    %=\rho\, \Omega^2 h \iint u_0^2(x,y) \rmd x\,\rmd y
    = \sigma\, h\iint \left(\left(\frac{\partial u_0}{\partial x}\right)^2+\left(\frac{\partial u_0}{\partial y}\right)^2 \right) \rmd x \,\rmd y,
    \hspace{0.5cm}
    \textnormal{and} \hspace{0.5cm}   m_{\mathrm{eff}}=\rho \,h \iint u_0^2(x,y)\, \rmd x \rmd y
\end{equation}  \label{Eqkmembrane}
expressed in geometric terms.

\begin{figure}
    \centering
    \includegraphics[width =178mm]{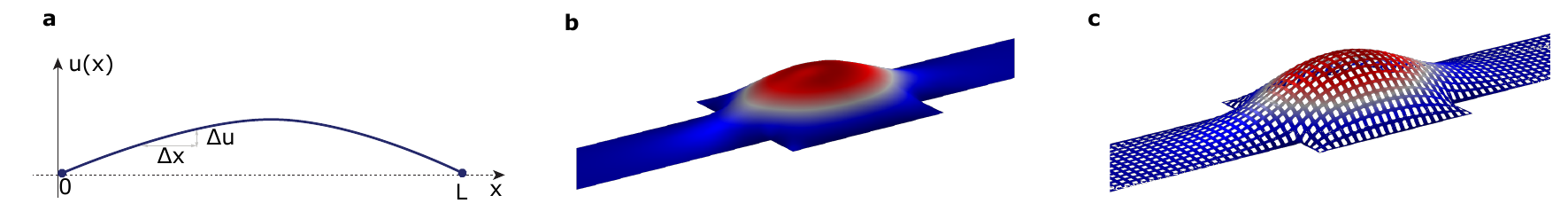}
    \caption{ (a) Elongation energy in a deformed string. (b) FEM simulation of a non-meshed acoustic fabry perot cavity. Estimated parameters: $k=418$ N.m$^{-1}$; $m_{\mathrm{eff}}=5.11\times10^{-13}$ kg; $\alpha=3.65\times10^{13}$ N.m$^{-3}$; $x_{\mathrm{crit}}=16.5$ nm;
$\Omega_{\textnormal{m}}=4.5$ MHz. (c) FEM simulation of a meshed acoustic Fabry Perot cavity. Estimated parameters: $k=253$ N.m$^{-1}$; $m_{\mathrm{eff}}=3.8\times10^{-13}$ kg; $\alpha=2.64\times10^{13}$ N.m$^{-3}$; $x_{\mathrm{crit}}=15$ nm; $\Omega_{\textnormal{m}}=4.1$ MHz. Physical parameters: $Y=250$  GPa; $\sigma=1$ GPa; $\rho=3100$ kg.m$^{-3}$, $Q=28 000$. Resonator width=length=80 $\mu$m, tunnel width 44 $\mu$m.}
    \label{figuresup1}
\end{figure}

Using the derived geometric terms, we can directly feed them into finite element models (FEM) such as COMSOL. The numerical results are calculated for simple systems such as strings or membranes and compared with the analytical expressions. Once verified, we modify the geometry and introduce the meshed membrane. The modeshapes obtained are plotted in Fig. \ref{figuresup1}b and \ref{figuresup1}c, from these simulation we can extract the non-linear coefficient for the meshed membrane $\alpha=2.64\times10^{13}$ N.m$^{-3}$.

\section{Electrostatic actuation and electric power requirements}

Sections \ref{sectioncriticalamplitudecriticalenergy} and \ref{energycostphonons} focussed on the acoustic energy required for nonlinear nanomechanical logic operation, considering acoustic inputs and outputs. Before the acoustic wave is present in the circuit, it is generated by electrostatic actuation. Here we discuss the parameters of this process, and discuss the overall (electric+acoustic) efficiency.

\subsection{Electrostatic actuation}

\begin{figure}[h!]
    \centering
    \includegraphics[width=178 mm]{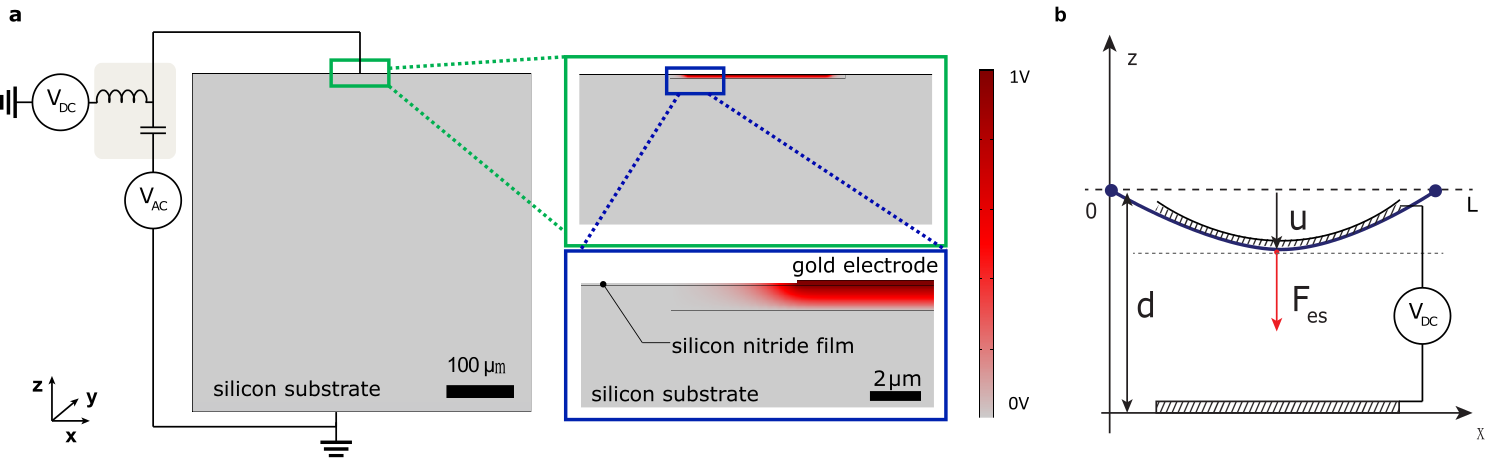}
    \caption{(a) 2D electrostatic simulation of the electric potential in response to a 1V DC bias applied between a gold electrode located above the released silicon nitride film and a ground plane below the silicon substrate. Because of the extremely large permittivity of doped silicon at MHz frequencies, the electric energy density is well localized in the undercut region between the silicon nitride and the silicon substrate. This means the capacitance can be well approximated by that of a parallel plate capacitor  ($C=\frac{\varepsilon_0 \varepsilon_r A}{d}$) with a plate separation distance $d$ equal to the depth of the undercut---with an error under 4\% here. Simulation parameters: substrate thickness 500 $\mu$m; undercut depth 1 $\mu$m; silicon permittivity $1 \times 10^4$; electrode width: 40 $\mu$m; simulated capacitance per unit length 367 pF/m; parallel plate approximation: 354 pF/m. (b) Schematic showing the relevant parameters for the estimation of stored mechanical and electrostatic energies.}
    \label{fig:electrostaticactuation}
\end{figure}

A finite element simulation of the electrostatic actuation process is shown in Fig. \ref{fig:electrostaticactuation}. It shows a cross-section of the 500 $\mu$m thick silicon wafer, covered by the 80~nm thick silicon nitride membrane, and 40 $\mu$m wide gold electrode. As these are $\sim$10 000 times thinner than the wafer, these are only visible in the zoomed-in sub-plots. Actuation of the membrane is provided by the electrostatic interaction between the gold electrode atop the phononic waveguide and an electrode on the underside of the wafer (see also Figs. 2 \& 3 of the main text). The silicon handle wafer is P-doped (Boron), with a resistivity of 1-10 Ohm cm. As such, its low frequency permittivity $\varepsilon$ is greatly enhanced over its optical frequency permittivity ($\varepsilon_{\mathrm{Si, THz}}=n^2=3.5^2\simeq 12$), due to charge transfer through the silicon, with a MHz-range permittivity of $\sim 10^4$ (so-called colossal permittivity \cite{liu_general_2018,sun_colossal_2019}). The effect of this is shown in Fig. \ref{fig:electrostaticactuation}, which plots the electric field in response to a 1V bias applied between both electrodes. Because of the extremely large permittivity of doped silicon at MHz frequencies, the electric energy density is well localized in the undercut region between the silicon nitride and the silicon substrate (see Fig. \ref{fig:electrostaticactuation}(c)). This means the capacitance can be well approximated by that of a parallel plate capacitor with an air gap ($C=\frac{\varepsilon_0  A}{d}$), with a plate separation distance $d$ equal to the depth of the undercut. The error from this approximation is under 4\% here, with a simulated capacitance per unit length (along $\hat{y}$)  of 367 pF/m from the finite element simulation and 354 pF/m from the parallel-plate approximation.

It is instructive to relate the mechanical energy stored in the membrane deflection, to the stored energy in the capacitor. As justified above, we employ the parallel plate approximation. The mechanical energy stored in the deflection of amplitude $u$ is:
\begin{equation}
    E_M=\frac{1}{2}k u^2=\frac{1}{2} k \left(\frac{\Fes}{k}\right)^2=\frac{1}{2}\frac{\Fes^2}{k}
\end{equation}
where $\Fes=\frac{1}{2}\frac{\varepsilon_0 A}{d^2} V^2$ is the electrostatic force applied on the membrane, in the limit of small displacements $u\ll d$. Similarly, the stored electrostatic energy in the capacitor is
\begin{equation}
    E_e=\frac{1}{2}C V^2 =\frac{1}{2} \frac{\varepsilon_0 A}{d} V^2=\Fes d
    \label{Eqelectricalenergyandforce}
\end{equation}
The ratio $\eta$ of these two energies, valid in the limit of small displacements, takes the form:
\begin{equation}
    \eta=\frac{E_M}{E_e}=\frac{\Fes}{2 k d}=\frac{\varepsilon_0 A V^2}{4 k d^3}=\frac{1}{2}\frac{u}{d}.
\end{equation}
Note that this energy analysis can be readily converted to a power analysis when the waveguide is operated sufficiently high above its cutoff frequency, such that the group velocity of the acoustic wave is close to its phase velocity: $v_g \sim vp \sim \sqrt{\frac{\sigma}{\rho}}$, in which case the acoustic energy has left the capacitor region before the start of a new cycle.
In the current experiments, where the capacitor is charged with an AC voltage source, there is no fundamental lower bound on power loss in the electric circuit (ohmic losses may in principle be made arbitrarily small). On the other hand, if operated in a regime where a given fraction of the capacitor energy is lost during the charge \& discharge process---as when the capacitor is charged from a constant voltage power supply, which is the case for CPUs, where this fraction reaches one half~\cite{zhirnov_minimum_2014}---it is beneficial to operate in a regime of small capacitor gap $d$, which maximizes the electrostatic force for a given capacitor energy, see Eq.(\ref{Eqelectricalenergyandforce}). Using an oxide sacrificial layer under the nitride as in Ref. \cite{baker_optical_2012}, this gap can reliably be made sub-100 nm.

\section{Critical amplitude and critical energy}
\label{sectioncriticalamplitudecriticalenergy}
Considering the total energy of the resonator $E$ as
\begin{equation}
\begin{aligned}
     E&=\frac{1}{2}k\,x^2+\frac{1}{4}\alpha\,x^4
     =
     \frac{1}{2} m_{\mathrm{eff}} \Omega_{\textnormal{m}}^2\left(1+\frac{\alpha}{2 k} x^2\right)x^2,
     %=\frac{1}{2}k\,x^2\left(1+\frac{\alpha}{2 k} x^2\right)\\
\end{aligned}
\end{equation}
where the resonance frequency becomes amplitude dependent
\begin{equation}
    \Omega
    =
    \Omega_{\textnormal{m}}\left(1+\frac{\alpha}{2k} x^2\right)^{\frac{1}{2}}.
\end{equation}
The frequency shift $\delta\Omega(x)$ is given by
\begin{equation}
\delta\Omega(x)=\frac{\alpha}{4k} \Omega_{\textnormal{m}} x^2, 
\end{equation}
where $\alpha$ is the non-linear coefficient. The critical amplitude is reached when this amplitude-dependent frequency shift $\delta \Omega \sim \Gamma=\Omega_{\textnormal{m}}/Q$, that is:
\begin{equation}
    x_{\mathrm{crit}}=\sqrt{\frac{2\, k\, \Gamma}{3\,\alpha\, \Omega_{\textnormal{m}}}}=\sqrt{\frac{2\, \Gamma\, \Omega_{\textnormal{m}}\, m_{\mathrm{eff}}}{3\, \alpha}}=\sqrt{\frac{2\, k }{3 \,\alpha \,Q}}.
    \label{Eqxcrit}
\end{equation}

Using these derivations, we calculate the linear elongation energy for a string \ref{figuresup1}(a), for an evanescently coupled membrane \ref{figuresup1}(b) and an evanescently coupled meshed membrane, as shown in \ref{figuresup1}(c). The meshed case has lower mass and frequency compared to the unperturbed membrane. The critical amplitude is however not significantly reduced, since both $k$ and $\alpha$ drop by comparable amounts (divided by 1.65 vs 1.4). This observation is expected, as the mode profile is essentially unperturbed by the subwavelength hole pattern.

Finally, the last relevant parameter is the critical energy $    E_{\mathrm{crit}}$, which is the energy cost associated to take the resonator amplitude to the critical amplitude, defined as 

\begin{equation}
    E_{\mathrm{crit}}=\frac{1}{2}k\,x_{\mathrm{crit}}^2=\frac{k^2}{3\, \alpha\, Q}=\frac{\meff^2\, \Omega_{\textnormal{m}}^4}{ 3\,\alpha \, Q}.
\end{equation}
It is important to notice that most of the energy stored in the resonator corresponds to the linear elastic energy. Since $\frac{1}{2}k x^2\gg \frac{1}{4}\alpha x^4$, the energy arguments presented in \eqref{eq:Elost} refer to the energy in the linear regime, but apply to the non-linear case.

\section{Energy cost derivation}
\label{energycostphonons}

In practice, the energy cost depends on other parameters, specifically the desired contrast between the `0' and `1' amplitudes, the duration of the logic pulses, and how much of the energy in the Duffing oscillator is transmitted into the output waveguide (and therefore not lost). We develop a simple model of the total energy cost, in the limit that the pulse duration is long compared to the decay time of the nanomechanical resonator and that the majority of energy in the gate comes from the pump. The model predicts different energy costs and functional dependencies for `0' and `1' output states. This can be understood directly from the difference in amplitudes of the two outputs: to achieve a lower amplitude for a `0' the gate must reject more of the input energy. The results of the model are shown as a function of the level of impedance matching of the resonator in Fig.~4{\bf d} of the main text using the calibrated Duffing coefficient and spring constant of our resonators.

\begin{figure}[h!]
     \centering
     \includegraphics[width= 80 mm]{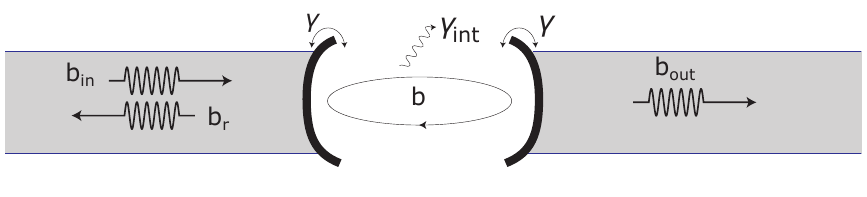}
     \caption{Schematic of the model used to calculate the energy cost of each logic operation. Double sided mechanical resonator excited by an input wave  of amplitude $x_{\rm{in}}$ with a coupling rate $\gamma$. The waves exiting the cavity on the right ($x_{\rm{out}}$) and left ($x_{\rm{r}}$) are also coupled out to the cavity at a rate $\gamma$. The intrinsic cavity loss rate is $\gamma_{\rm{int}}$. }
     \label{fig:cavityScheme}
\end{figure}

To estimate the energy cost for a single operation in our device, we consider the double-sided non-linear mechanical cavity of Figure~\ref{fig:cavityScheme} which oscillates with an amplitude proportional to  $|b|$ with an energy $E = |b|^2$, when excited by an input acoustic wave of power $|b_{\rm{in}}|^2$. The power of the wave exiting the right side of the cavity is $|b_{out}|^2$ and the power of the wave reflected by the cavity is $|b_{r}|^2$. 
The equation of motion of such cavity is given by~\cite{aspelmeyer_cavity_2014}:
\begin{equation}
    \dot{b} = -\frac{\gamma_{\rm{tot}}}{2}b-i\Delta b +\sqrt{\gamma}b_{\rm{in}} 
\end{equation}
where $\gamma_{\rm{tot}} = 2 \gamma + \gamma_{\rm{int}}$ is the total energy 
loss rate, $\gamma$ the coupling rate of both sides of the cavity, $\gamma_{\rm{int}}$ the intrinsic loss rate of the cavity, $\Omega$ the angular frequency of the exciting wave, $\Delta = \Omega -\Omega_{\rm{m}}$ the 
drive frequency detuning from the cavity resonance $\Omega_{\rm{m}}$.
This equation can be solved for the steady state when  the duration of the acoustic wave pulse $T=N/\gamma_{\rm{tot}}$ is larger than the inverse of the decay rate. This is the case here since the quality factor of the resonators is around $10^4$ leading to a decay rate of hundreds of Hertz which inverse is two orders of magnitude smaller than the pulse duration used. 
In this situation, $N \gg 1$ and $\dot{b} \sim 0$ which leads to:
\begin{equation}
    b = \frac{\sqrt{\gamma}}{\gamma_{\rm{tot}}/2+i\Delta} b_{\rm{in}}.
    \label{eq:b}
\end{equation}
%% Efficiency
As shown by figure~\ref{fig:cavityScheme}, $b_{\rm{out}} = \sqrt{\gamma}b$, therefore the steady-state efficiency defined as $\eta = |b_{\rm{out}}|^2/|b_{\rm{in}}|^2$ can be expressed as
\begin{equation}
    \eta = \frac{\gamma^2}{ \gamma_{\rm{tot}}^2/4 +\Delta^2}
    \label{eq:eta}
\end{equation}

Using eq.~\ref{eq:b}, the mechanical energy of the cavity $E$ is then:
\begin{equation}
    E = \frac{\gamma}{\gamma_{\rm{tot}}^2/4+\Delta^2} |b_{\rm{in}}|^2.
\end{equation}
The steady-state energy can be expressed as a fraction $A$ of the critical energy $\Ecrit$, $E = A\Ecrit$, where $A = A^{(1)} >1$ when the duffing oscillator is in the high amplitude state corresponding to a logic 
operation output ``1'' and $A = A^{(0)} <1$ when the oscillator is in the low amplitude state, corresponding to a logic operation output  ``0''. 
Therefore we have:
\begin{equation}
    |b_{\rm{in}}|^2 = \frac{A \Ecrit (\gamma_{\rm{tot}}^2/4+\Delta^2)}{\gamma}.
    \label{eq:bin}
\end{equation}
Since the right output of the cavity guides the result of the logic operation, only the wave propagating away from the cavity in the left direction and the energy lost to the environment at a rate $\gamma_{\rm{int}}$  
will contribute to the losses (see figure \ref{fig:cavityScheme}). Therefore the energy lost $\Elost$ during 
the logic operation can be expressed as~\cite{aspelmeyer_cavity_2014}
\begin{equation}
    \Elost = \int_{0}^{T} (|b_{\rm{in}}|^2 - |b_{\rm{out}}|^2) \,dt.
\end{equation}
Since $b_{\rm{out}} = \sqrt{\gamma}b$ and, for the steady state, $b_{\rm{in}}$ can be assumed to be constant as function of time. Therefore using eq. \ref{eq:b} and \ref{eq:bin},
\begin{equation}
    \Elost =\frac{ N A  \Ecrit}{\gamma_{\rm{tot}}\gamma} \left(\frac{\gamma_{\rm{tot}}^2}{4} + \Delta^2 - \gamma^2\right)
    \label{eq:Elost}
\end{equation}

When the output of the logic operation is a binary ``1'', $A = A^{(1)} >1$ and $\Delta = \Delta^{(1)} =0$ 
since the cavity has shifted on resonance. Using eq.~\ref{eq:Elost}, the energy lost for an output ``1'' (Output 1 in Fig. 4d in the manin text),
 $\Elost^{(1)}$ is therefore:
 \begin{equation}
    \boxed{\Elost^{(1)} =  \frac{ N A^{(1)} E_{\rm crit}}{4} \left (\frac{\gamma_{\rm int}}{\gamma} \right ) \left(2- \frac{\gamma_{\rm{int}}}{\gamma_{\rm tot}} \right) }
    \label{energy1}
\end{equation}
where we have used the relation $\gamma_{\rm tot} = 2\gamma + \gamma_{\rm int}$.

When the output of the logic is a binary ``0'', $A = A^{(0)} <1$. However, the detuning is not negligible since the resonance frequency of the duffing oscillator has not shifted to the driving frequency. Knowing that the acoustic pump signal excites the oscillator close to the critical energy independently if the logic operation result is a ``0'' or a ``1'', the contribution of the acoustic signal A and B to the input acoustic power will be negligible and we can assume that $b_{\rm{in}}^{(1)} \sim b_{\rm{in}}^{(0)}$. The detuning can then be expressed using equation~\ref{eq:bin} as :
\begin{equation}
    \Delta^{(0)} = \sqrt{ \frac{\gamma_{\rm{tot}}^2}{4}\left(\frac{A^{(1)}}{A^{(0)}}-1\right)}.
\end{equation}
Using equation~\ref{eq:Elost} it is straightforward to see that the energy lost when the output of the logic operation is a ''0`` is (Output 0 in Fig. 4d in the main text):
\begin{equation}
    \boxed{\Elost^{(0)} = \frac{N A^{(1)} E_{\rm crit}}{4} \left( \frac{\gamma_{\rm{tot}}}{\gamma } - \frac{4 \gamma}{\gamma_{\rm{tot}} } \left ( \frac{ A^{(0)}}{A^{(1)}} \right ) \right) }
    \label{energy0}
\end{equation}
The evolution of $\Elost^{(0)}$ and $\Elost^{(1)}$ as function of $Q/Q_{\rm{int}} = \gamma_{\rm{int}}/\gamma_{\rm{tot}}$ is displayed in the main text figure 4d). For this figure, we use the experimental values of $\Omega_{\textnormal{m}}$, $\gamma_{\rm{int}}$, $\gamma$, $\alpha$ and $N$. $\meff$ is calculated from finite element simulation using the equations of section \ref{sec:DerivationDuffing}.

\begin{table}[h!]
    \centering
    \begin{tabular}{|l| c|c|c|}
         \hline
       \rule{0pt}{2ex} \textbf{Parameter}  & \hspace{1mm}\textbf{
       Energy value (fJ)}  \hspace{1mm}&  \textbf{
       Energy value (fJ)} &\textbf{Expression} \hspace{1mm}\\
     & (Gate duration 10 ms)& (Gate duration 120 ms) &\\
 \hline
       Energy lost (0)& 38 &455 &$\displaystyle \Elost^{(0)} = \frac{N A^{(1)} E_{\rm crit}}{4} \left( \frac{\gamma_{\rm{tot}}}{\gamma } - \frac{4 \gamma}{\gamma_{\rm{tot}} } \left ( \frac{ A^{(0)}}{A^{(1)}} \right ) \right)$\\ 
       \hline
       \rule{0pt}{4ex} Energy lost (1)  & 7 &89&$\displaystyle \Elost^{(1)} =  \frac{ N A^{(1)} E_{\rm crit}}{4} \left (\frac{\gamma_{\rm int}}{\gamma} \right ) \left(2- \frac{\gamma_{\rm{int}}}{\gamma_{\rm tot}} \right) $  \\ 
       \hline
       \rule{0pt}{4ex} Critical energy&  28 &28 & $E_{\mathrm{crit}}=\displaystyle\frac{1}{2}k\,x_{\mathrm{crit}}^2=\frac{k^2}{3\, \alpha\, Q}=\frac{\meff^2\, \Omega_{\textnormal{m}}^4}{ 3\,\alpha \, Q}$\\ \hline  
         \rule{0pt}{4ex} Average energy cost &22&272 & $\bar{E}=\displaystyle\frac{1}{2}\left(\Elost^{(1)}+\Elost^{(0)}\right)$\\ 
         \hline
        \rule{0pt}{4ex} Minimum average energy cost& 12 &140 &$\displaystyle 0=\frac{d \bar{E}(Q)}{d Q}$\\  \hline
    \end{tabular}
    \caption{Values of the energies involved in the gate operation considering a $Q=28000$.}
    \label{tableEnergies}
\end{table}

As can be seen, the energy cost for an output `1' state increases monotonically as the resonator becomes increasingly undercoupled, while going to a constant minimum value of $E_{\rm lost, min}^{(1)} = N A E_{\rm crit} (\gamma_{\rm int}/\gamma_{\rm tot})$ in the impedance-matched regime where the loaded $Q$ is much smaller than $Q_{\rm int}$. By contrast, the energy cost for an output `0' state is optimal when $Q/Q_{\rm int}=0.49$, in between the well- and poorly-impedance matched regimes. The increased energy cost at low $Q$ in this case can be understood because as the $Q$ decreases the decay rate of the resonator increases, so that a larger nonlinear frequency shift is required to operate the gate. Achieving this larger shift requires a higher amplitude of drive, and therefore increases the energy cost.

Our experiments operate in the near-impedance-matched  regime, with $\gamma_{\rm int}/\gamma \sim 0.2$, $\gamma_{\rm int}/\gamma_{\rm tot} \sim  0.1$, and $\gamma_{\rm tot}/\gamma \sim 1$. The measured oscillation amplitudes of 2~nm and 18~nm, respectively, for a `0' and a `1' give a ratio $A^{(0)}/A^{(1)} = (2/18)^2 \sim 0.01$. The oscillation energy of a `1' is around forty percent higher than the critical energy, with $A^{(1)} = (x^{(1)}/x_{\rm crit})^2 = (18 \, {\rm nm}/15 \, {\rm nm})^2 = 1.4$. For the data shown in the main text, $N \sim 18$, while for our shorter 10~ms duration gates, this reduces to $N \sim 1.4$. Using these values in Eqs.~(\ref{energy0})~and~(\ref{energy1}), we find that the energy lost during an operation of our longer gates is roughly $2.4 E_{\rm crit}$ and $14 E_{\rm crit}$ for an output `1' and `0', respectively; while the equivalent energy losses for our shorter gates are roughly $0.25 E_{\rm crit}$ and $1.35 E_{\rm crit}$.

\section{Drift correction}

Each point $i$ of the four traces shown in Fig. 5 b represents the mean value around the center (withing the resolution bandwidth window) of each NAND gate. During the data acquisition of error statistics (main text figure 5 b)), a slow drift of the amplitude of each window was observed, due to thermal effects. 
The maximum drift was observed for the red trace which dropped by 2.7~dBm after 0.6~hours of continuous measurements. To remove this systematic error from our measurement the drift was compensated by applying the same simple linear transformation to the four traces. That is to say, each point $i$ of the four traces is transformed $i\rightarrow i'$ as $i' = m \times i+p$. The blue trace (named $A$) and red trace (named $B$) of figure~5 b) were used as a reference. We define the variables $m$ and $p$ as:
\begin{equation}
    m = \frac{\overline{A}-p}{\langle A \rangle},\hspace{1 cm}     p = \frac{\overline{B}\langle A \rangle-\overline{A}\langle B \rangle}{\langle A \rangle - \langle B \rangle},
\end{equation}
where $\overline{X}$ represents the average of $X$ over the 1500 measurements and $\langle X \rangle$ represents the moving average of $X$ over a window of 100 measurements centered on point $i$ but excluding point $i$. The resulting corrected data is presented in Fig.~5 b showing that the systematic error caused by drift is corrected efficiently. Histograms are plotted on the right of Fig. 5 b. These are found to be consistent with Gaussian distributions (solid lines), as expected from the central limit theorem.

\section{Five requirements of a logic device}
Five requirements have been identified as necessary for a viable logic device \cite{waser_nanoelectronics_2012,behin-aein_proposal_2010}: nonlinearity, gain, cascadability, feedback prevention and functional completeness. A device that meets these requirements is viable in the sense that it is effective and scalable. Previous nanomechanical devices have achieved strict subsets of these requirements, but not met all five simultaneously. We posit that our proposed device is the first nanomechanical system that does meet meets all five criteria simultaneously.

 \begin{figure}[h!]
     \centering
     \includegraphics[width= 88mm]{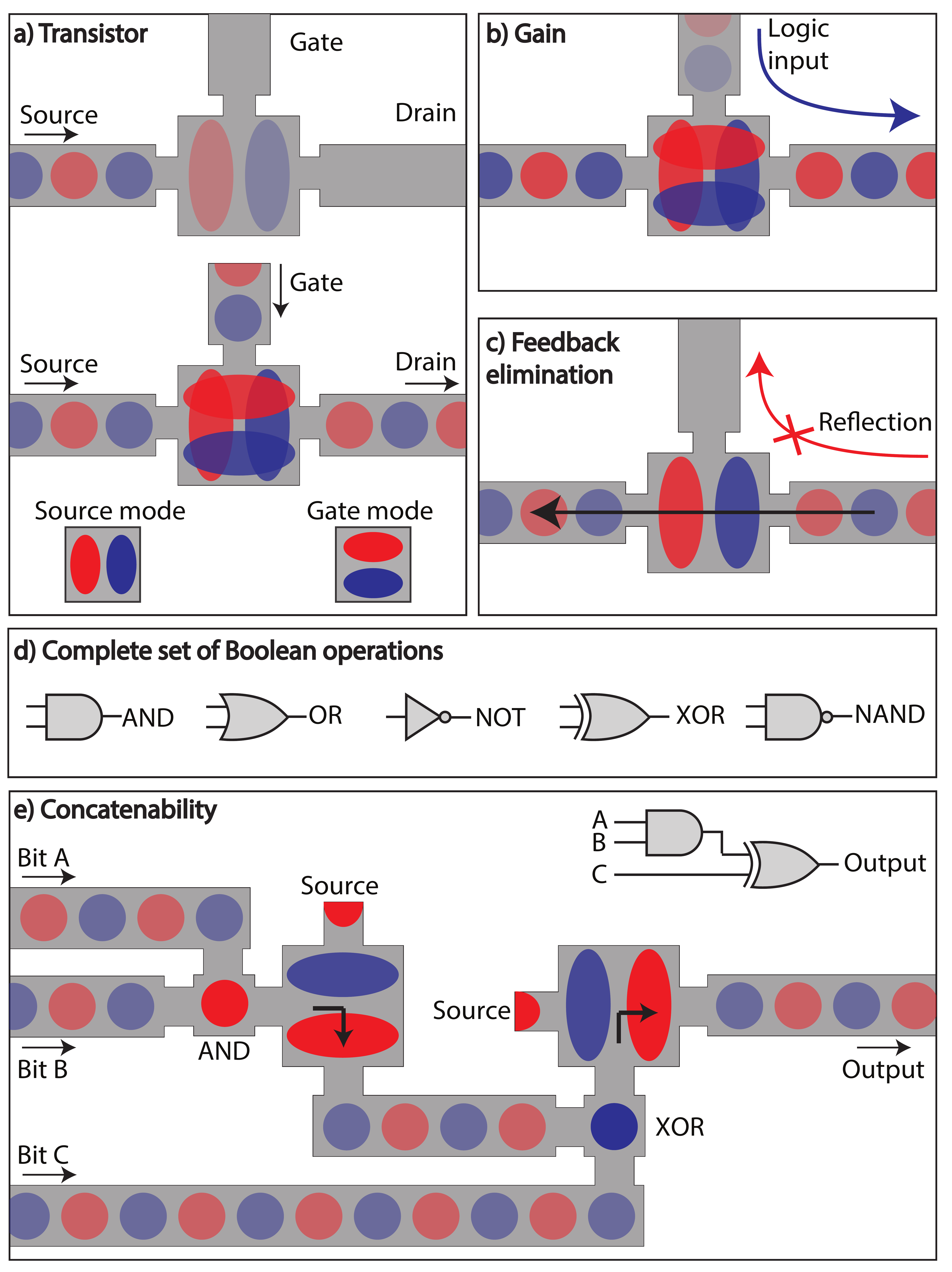}
     \caption{FIVE-essentials: a) Diagram of a nanomechanical transistor used for feedback prevention. The source, gate and drain are formed by input, and output waveguides. The time domain simulations are shown in Fig.~\ref{fig:feedbackprevention}(a). b) Configuration of a gate for gain $\geq 1$, where the pump (left waveguide) provides majority of the oscillation energy. The logic input, carrying a small fraction of the energy is delivered from the top waveguide. The logic output of the gate is sent into the next gate through the right waveguide. c) In the transistor shown in a) the feedback is eliminated using using orthogonal but degenerate modes between the between the logic input and the logic output. Further simulations are shown in Figure~\ref{fig:feedbackprevention}~(b)-(c). The complete set of Boolean operations is composed the gates AND, OR, NOT, XOR and NAND. e) Schematic of cascadability architecture for our devices where the ultrasonic outputs waves are carrier in the mode as the input, allowing multiple devices to concatenate in the same circuit.}
     \label{fig:FIVEessentials}
\end{figure}

\begin{figure}[h!]
    \centering
    \includegraphics[width=88mm]{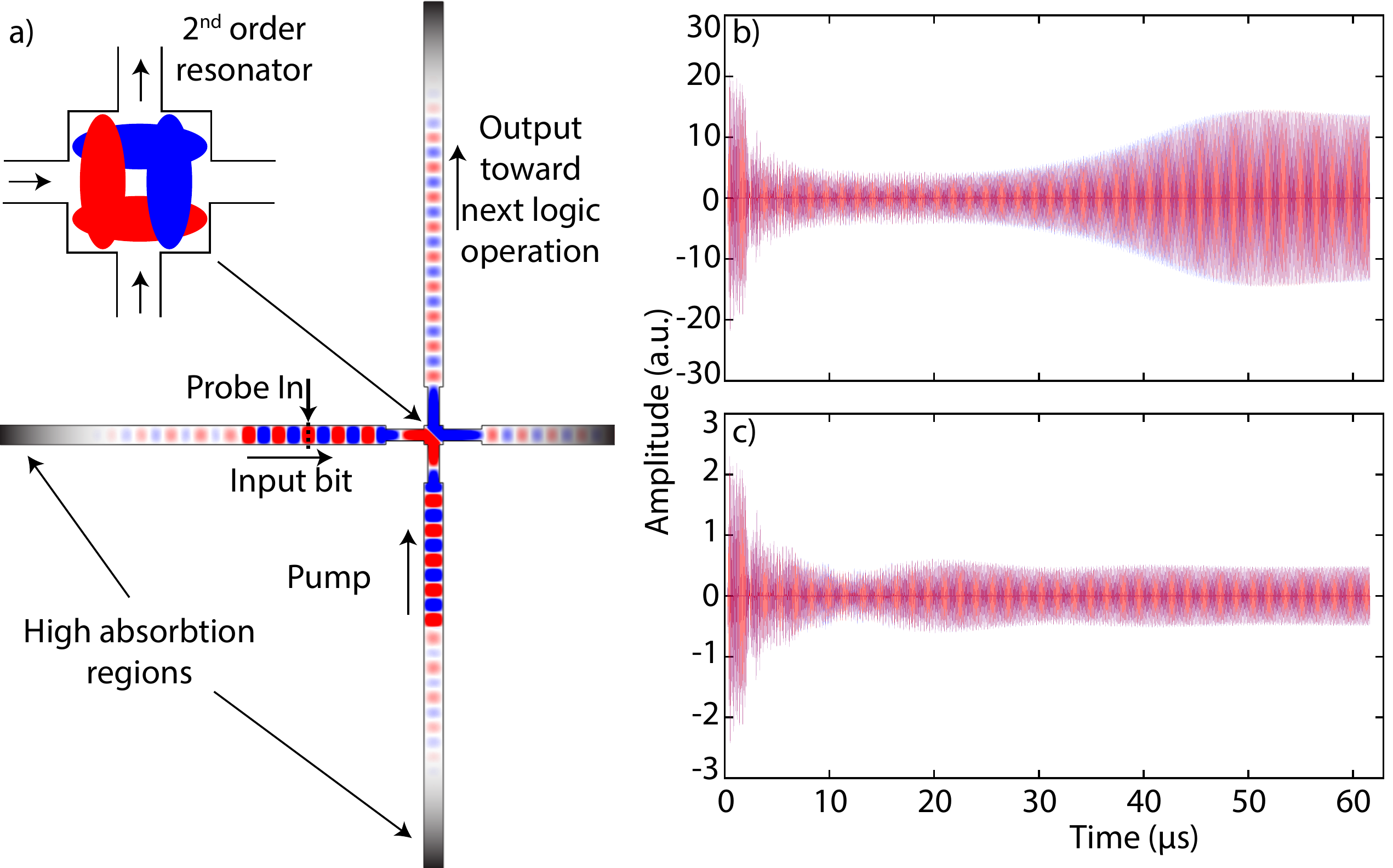}
    \caption{(a) Diagram of the transistor mechanism, where the source, drain and gate signals are provided by the bottom, top, and left waveguides respectively. Each waveguide contains absorbing regions that were toggled on and off in the simulation. (b) Amplitude in gate waveguide when pumping the resonator and inputting a `1' (high amplitude) gate signal. The plot contains two time series drawn as red and blue lines, which correspond to running the simulation with and without absorption in the drain waveguide. The series are identical which shows reflections are not affecting the gate waveguide. (c) As previous but with a `0' (low amplitude) gate signal.
    \label{fig:feedbackprevention}}
\end{figure}

\begin{enumerate}
\item Non-linear characteristics (for signal-to-noise)

Nonlinearity is necessary to achieve a good signal-to-noise ratio \cite{nikonov_simulation_2007}. A linear logic gate would poorly distinguish between `1' and `0' responses. Ideally, a gate should show a marked change in output when the input crosses from the `1' to the `0' preimage, and vice versa. The more easily a `1' can be told apart from a `0', the less noise is introduced into the logical circuit by the gate. Figure~4~(c) in the main text shows the clear difference in the two amplitudes of oscillation introduced by Duffing bistability in our resonators. The probability of mislabelling the amplitudes is negliglbe, as seen in Figure~5~(c).

\item Power amplification

To string multiple logic gates together without gradually diminishing the signal, the gain of a single gate must be $\geq 1$. In our gate this is achieved by the pump signal, which provides the majority of the oscillation energy in each logic operation. The pump allows an input signal with only a fraction (for example, 10\%) of the jump up energy to induce a jump up. The output signal from the pump-driven resonator is more energetic than this fraction---i.e. gain is $>1$---more than making up for any propagation losses that will occur before the next resonator. 

\item Cascadability

Logic circuits with complex truth tables typically require multiple logic devices. Additionally, inputs and outputs of gates may come from and go into other, non-logic components. Therefore a useful logic device should produce an output compatible with other components in the circuit, including other gates. This is a requirement that has not been met by previous NEMS work, usually for one of two reasons: (i) the gate is bookended by a pair of electrical actuators, which have unscalable efficiencies on the order of $10^{-7}$ \cite{wenzler_nanomechanical_2014}; or (ii) the gate outputs a mechanical signal, but in a different mode which cannot serve as input for another gate \cite{hatanaka_broadband_2017}. In contrast our device outputs ultrasonic waves in the same mode as the input, allowing multiple devices to concatenate in the same circuit.

\item Feedback prevention (output does not affect input)

To sequence multiple logic gates, the output channel of a gate should not affect either of its input channels. Our logic gate in its simplest form does not have this property, because energy from the resonator leaks into both the input and output waveguides. However, feedback prevention can be achieved by pairing the gate with an `acoustic transistor' device that we have designed. We propose using transistors immediately logic gates, as illustrated in Figure~6~e), prevent unwanted feedback in a logic circuit.

The transistor device has been simulated with a finite difference time domain code \cite{mauranyapin_tunneling_2021}. Figure~\ref{fig:feedbackprevention}~(a) provides a diagram of the device and the pump, input, and output waveguides. The pump is blue detuned and therefore by default transmits inefficiently across the resonator to the output waveguide. Efficient transmission is enabled by the gate signal, which increases the resonance frequency of the resonator through Duffing stiffening. The gate signal thereby determines the output signal. Feedback in the opposite direction is prevented because the gate and pump signals induce orthogonal modes in the resonator, which do not couple. This was tested in simulation by allowing and disallowing reflections in the output waveguide: Figure~\ref{fig:feedbackprevention}~(b)-(c) show that the reflections did not change the amplitude in the input waveguide.

\item Complete set of Boolean operations

A logic device should be functionally complete, that is, able to construct any desired truth table. Functional completeness is certainly provided by the set of all standard logic gates (AND, NOT, OR, and so on), however it can also be provided by smaller subsets. In particular, sequences of NAND gates can reproduce the truth tables of any other gate \cite{natarajan_fundamentals_2020}. Therefore by achieving a NAND gate we have shown that Boolean completeness may be achieved within our nanomechanical architecture. We also note that other logic gates can be achieved directly without sequencing NAND gates---for example, we experimentally achieved OR and AND gates.

\end{enumerate}

\section{Photocurrent derivation}
%Show what is the photocurrent expression as function of the local oscillator and signal field amplitude. This is to explain the calibration method.
To measure the displacement of the meshed membrane, we use an optical heterodyne detection system. In this measuring system, the optical probe field $E_p$ is reflected off the oscillating membrane (see main text figure 3). For a given polarisation, the probe field can be expressed as:
\begin{equation}
    E_{\textnormal{p}} = A_{\textnormal{p}} \exp(i\omega t + \phi),
\end{equation}
where $A_{\textnormal{p}}$ is the amplitude of the field, $\Omega$ its frequency, $\phi$ is the phase modulation created by the change of the light path length due to the membrane motion. Here for simplicity, we have neglected any noise in amplitude and phase of the field and we have chosen a fixed observation position at the position of the detector. Similarly, the local oscillator field $E_{LO}$ can be expressed as:
\begin{equation}
    E_{\textnormal{LO}} = A_{\textnormal{LO}} \exp\left[i(\omega + \omega_{\textnormal{AOM}} )t \right]
\end{equation}
where $A_{\textnormal{LO}}$ is the amplitude of the field and $\omega_{\textnormal{AOM}}$ is the frequency shift created by the acousto optical modulator (AOM). These two fields are interfered and once detected, the photocurrent $i$ transduced by the detector is proportional to:
\begin{equation}
    i \propto A_{\textnormal{p}}^2 + A_{\textnormal{LO}}^2+ 2 A_{\textnormal{p}} A_{\textnormal{LO}} \cos(\omega_{\textnormal{AOM}}t-\phi ). 
\end{equation}

Since the membrane is driven with a sinusoidal excitation at frequency $\omega$, $\phi$ can be expressed as $\phi = A_\phi \cos(\Omega t )$ with $A_\phi$ the amplitude of the phase modulation in radian. Using the Jacobi-Anger relation, to first order, the response of the detector at the frequency $\omega_{\textnormal{AOM}} \pm \Omega$ is:
\begin{equation}
    i(\omega_{\textnormal{AOM}} \pm \Omega) \propto -\frac{1}{2} A_{\textnormal{p}} A_{\textnormal{LO}} A_\phi \left[ \sin (\omega_{\textnormal{AOM}}-\Omega)t\right) - \sin\left((\omega_{\textnormal{AOM}}+\Omega_{\textnormal{m}})t)\right],
\end{equation}
with the amplitude of the photocurrent at the frequencies $\omega_{\textnormal{AOM}} \pm \Omega$ is proportional to the amplitude of the membrane motion. Meanwhile the photocurrent at frequency $\omega_{\textnormal{AOM}}$ is given by
\begin{equation}
    i(\omega_{\textnormal{AOM}}) \propto A_{\textnormal{p}} A_{\textnormal{LO}} \cos(\omega_{\textnormal{AOM}}t).
\end{equation}

 Furthermore, to first order the ratio $i(\omega_{\textnormal{AOM}}\pm \Omega)/i(\omega_{\textnormal{AOM}})=A_{\phi}/2$. Knowing that $A_{\phi}=2\pi$ corresponds to 780~nm, the free space wavelength of the laser, we can directly retrieve the amplitude of motion of the membrane in nanometers. Typically, in our experiment, when the membrane is driven to high amplitude, the amplitude of the photocurrent at $\omega_{\textnormal{AOM}}$ and at $\omega_{\textnormal{AOM}} \pm \Omega$ is 0.14~V$_{\rm{RMS}}$ (-4~dBm) and 0.022~V$_{\rm{RMS}}$ (-20~dBm) respectively which give an amplitude of motion of $\sim$19~nm.

\section{Scaling with resonator size}

Here we quantify the gate energy cost scaling with size of a resonator. We consider an out-of-plane flexural motion of a square membrane of side $L$, thickness $h$, density $\rho$, Youngs's modulus $Y$, under tensile stress $\sigma$. The mode profile $u$ takes the form:
\begin{equation}
    u(x,y)=\sin\left( \frac{n \pi x}{L}\right) \sin\left( \frac{m \pi y}{L}\right).
\end{equation}
We consider hereafter the fundamental $n=1; m=1$ mode, with effective mass given by Eq.\eqref{eq:meffmembrane}:
\begin{equation}\label{eq:meffmembrane}
    m_{\mathrm{eff}}=\frac{1}{4} L^2 h \rho,
\end{equation}
its spring constant, given by Eq.\eqref{eq:kmembrane}, is:
\begin{equation}\label{eq:kmembrane}
    k=\frac{1}{2} \pi^2\, h\, \sigma,
\end{equation}
and the nonlinear Duffing coefficient given by Eq.\eqref{eq:alphamembrane} is:
\begin{equation}\label{eq:alphaduffingmembrane2}
    \alpha=\frac{9 \pi^4 \,h\, Y}{64\, L^2}.
\end{equation}

In the linear regime, the oscillation has a small-amplitude, the resonance frequency $\Omega_{\textnormal{m}}$ is:
\begin{equation}
    \Omega_{\textnormal{m}}=\sqrt{\frac{k}{m_{\mathrm{eff}}}}=\frac{\sqrt{2}\,\pi}{L}\sqrt{\frac{\sigma}{\rho}},
    \label{Eqomegamembrane}
\end{equation}
where we recognize the speed of sound $\sqrt{\frac{\sigma}{\rho}}$.
The critical amplitude $x_{\mathrm{crit}}$ given by Eq.\eqref{Eqxcrit} is:
\begin{equation}
    x_{\mathrm{crit}}=\frac{8 L}{3 \sqrt{3} \pi}\sqrt{\frac{\sigma}{Q Y}},
    \label{Eqxcritmembrane}
\end{equation}
with $Q$ the mechanical quality factor. The critical amplitude is linearly proportional to the resonator size $L$. The critical energy $E_{\mathrm{crit}}$ associated with reaching this critical amplitude is given by:
\begin{equation}
\boxed{
    E_{\mathrm{crit}}=\frac{1}{2}k\,x_{\mathrm{crit}}^2=\frac{16 L^2 h\, \sigma^2}{27 Q Y}.}
    \label{EqEcritmembranewithsize}
\end{equation}

This energy scales with the resonator area $L^2$ and thickness $h$, underscoring the merits of miniaturization. A large Young's modulus is also beneficial, as it increases the Duffing nonlinearity (which arises in response to an elongation of the material - see Eq. \eqref{eq:alphamembrane}) and therefore reduces the critical energy. Reducing the tensile stress $\sigma$ also reduces the critical energy, at a cost of a reduced operational frequency and reduced impedance mismatch between the suspended membrane and the substrate, leading to increased acoustic radiative losses. 

As an example, here we estimate the prospects for miniaturization of our platform considering a resonator made of a membrane with thickness as low as 10~nm and tensile stress of $\sigma\sim 10$~MPa.

We use Eq. \eqref{EqEcritmembranewithsize} to estimate the potential energy cost reduction for nonlinear nanomechanical logic. A significant step forward towards the can be achieved by using a resonator made of a Si$_3$N$_4$ membrane with thickness of 10~nm and tensile stress of $\sigma\sim 10$~MPa and a $Q$ factor of $\sim 4000$. With these feasible parameters, our platform would operate at the Landauer limit, having a resonance frequency of $\sim 180$~MHz and speed of gate operations $\sim 50$~kHz. The ultimate reduction in thickness can be achieved through the use of a purely two-dimensional nanomechanical resonator, made from a 2D material such as graphene
\cite{bunch_electromechanical_2007, lee_measurement_2008, lee_estimation_2012,barton_photothermal_2012,cole_evanescent-field_2015, inoue_resonance_2017,zhang_coherent_2020} or molybdenum disulfide (MoS$_2$)~\cite{castellanos-gomez_single-layer_2013}. Such materials have been employed to build high-Q suspended membrane resonators exhibiting hardening Duffing nonlinearity~\cite{inoue_resonance_2017, castellanos-gomez_single-layer_2013}, and are therefore compatible with our scalable approach to nanomechanical computing.

We consider a square single-layer graphene drum resonator operating at the Landauer limit with critical energy $\Ecrit=\log(2) k_{B} T$. The required parameters for this resonator are shaded in Table \ref{tablegrapheneresonator} and compared with our current platform.

\begin{table}[h!]
    \centering
    \begin{tabular}{|l| c|c|c|c|}
         \hline
       \textbf{Parameter}  &  \hspace{1mm}\textbf{Symbol}  \hspace{1mm}&  \textbf{
       Meshed Si$_4$N$_3$ Membrane} &\textbf{
       Graphene Membrane}& \hspace{1mm}\textbf{Unit} \hspace{1mm}\\
         \hline
 \rowcolor{lightgray}
 Young's modulus & Y &250& 1000 & GPa\\
          \rowcolor{lightgray}
          Size & $L$ & 80& 0.1 & $\mu$m\\
           \rowcolor{lightgray}
           Quality factor & $Q$ & 28000& 60 & NA\\
       Resonance frequency & $\Omega_{\textnormal{m}}/2\pi$ &4.1& 20000 & MHz \\
     Duffing coefficient &  $\alpha$ & 2.6 $\times 10^{13}$ &$4.1 \times 10^{17}$ & N/m$^{3}$\\
     
        Critical amplitude & $\xcrit$ &  $15 \times 10^{-9}$&  $1 \times 10^{-10}$ & m \\
        
        Critical energy & $ \Ecrit$ & $2.6 \times 10^{-14}$  & $1.23 \times 10^{-21}$& J \\
         Critical energy & $ \Ecrit$ & $7.2\times 10^6$ & $\log(2)$& $k_{\textnormal{B}}\, T$\\
\hline
    \end{tabular}
    \caption{The parameters in the shaded region for Si$_4$N$_3$ are determined from our current experimental system. The parameters in the shaded region for graphene membrane are chosen as feasible experimental values for platform miniaturization. The rest of the parameters are derived here from Eqs. \eqref{eq:alphaduffingmembrane2}, \eqref{Eqxcritmembrane} and \eqref{EqEcritmembranewithsize}. These parameters are represented as dashed lines in Fig.~\ref{fig:Miniaturization} calculated assuming a square membrane resonator. These values are in good agreement with those derived from FEM simulations. The small differences arise due to the presence of the tunnel barriers which modify the mode shape, which is not taken into account here.}
    \label{tablegrapheneresonator}
\end{table}

\begin{figure}[h!]
     \centering
     \includegraphics[width= 178 mm]{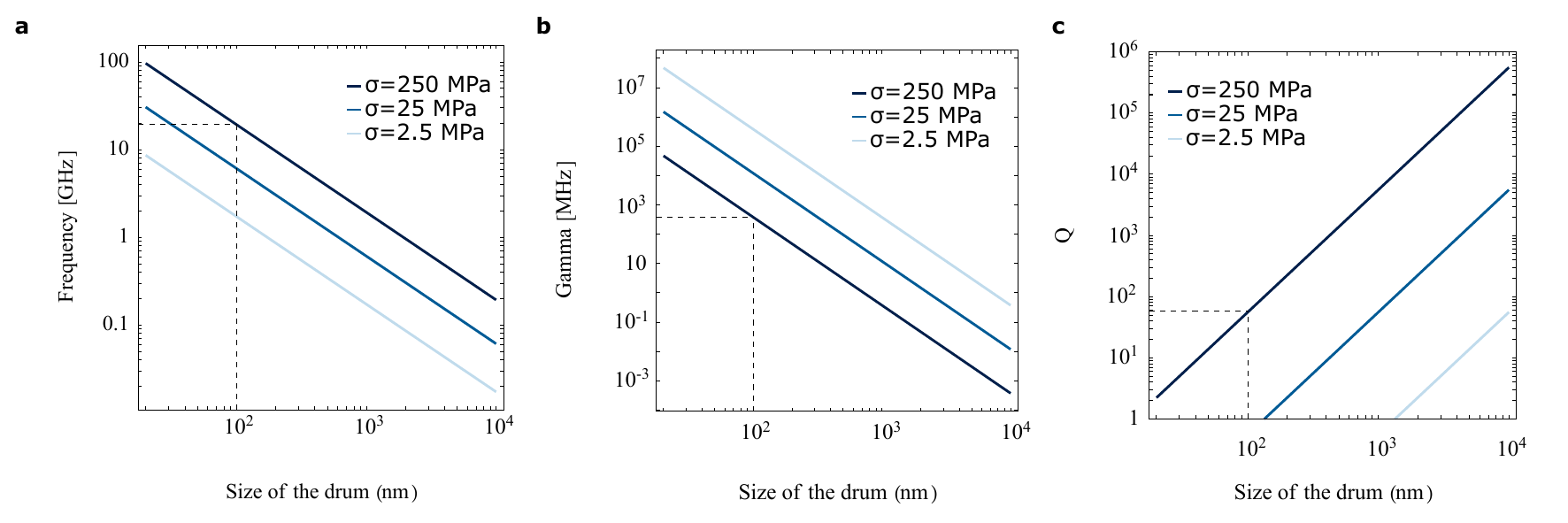}
     \caption{A graphene drum resonator operating at the Landauer limit for three cases of intrinsic stress $\sigma= 2.5$, 25 and 250 MPa. a) Operating frequency of a graphene drum as a function of the drum size. b) Dissipation rate of the drum resonator, therefore maximum gate speed. c) Lower bound to operate at the Landauer limit for the quality factor of the graphene drum as a function of the drum size.}
     \label{fig:Miniaturization}
\end{figure}

 In Fig.\ref{fig:Miniaturization} we plot the predicted operating frequency (Fig.\ref{fig:Miniaturization}a), speed of gate operations given by the dissipation rate (Fig.\ref{fig:Miniaturization}b) and the minimum required quality factor (Fig.\ref{fig:Miniaturization}c), all for a nanomechanical logic gate based on a single layer graphene membrane. These predictions are calculated for three different intrinsic stresses and as a function of the size of the graphene drum. For the case where the drum size is $L=100~$nm, Eqs. (\eqref{eq:meffmembrane}, \eqref{Eqomegamembrane}, \eqref{EqEcritmembranewithsize}) predict a resonator with a resonance frequency of $\sim20$ GHz and gate speed in the range $\sim$100 MHz. These calculations were done using a Young's modulus for graphene on the order of 1 TPa~\cite{bunch_electromechanical_2007,inoue_resonance_2017,lee_measurement_2008,lee_estimation_2012}, a density $\rho_{\mathrm{2D}}=8\times 10^{-19}$ kg/$\mu$m$^2$~\cite{barton_photothermal_2012, cole_evanescent-field_2015} for single-layer graphene, a thickness $h=0.3$ nm and tensile stress up to $\sigma=250\times 10^6$ Pa (consistent with the literature \cite{lopez-polin_increasing_2015,fan_graphene_2019}).

All parameters are summarized in Table \ref{tablegrapheneresonator}. This change in resonator design leads to a significant reduction in $\Ecrit$ of the order of 7 million, down to a value of only $\ln 2  k_B T$ reaching the Landauer limit for irreversible computing~\cite{landauer_irreversibility_1961}.


%\bibliography{supp_ref.bib}

%\bibliography{Nanomechanical_gate.bib}

\end{document}